\documentclass[aps,prx,twocolumn,longbibliography,eqsecnum,superscriptaddress]{revtex4-2}
\usepackage{amsmath,amsfonts, amssymb}
\usepackage{bm}
\usepackage{graphicx}
\usepackage{sidecap}
\usepackage{verbatim}
\usepackage{hyperref}
\usepackage{xcolor}
\usepackage{enumerate}
\usepackage{tikz,cleveref}
\usepackage{bbold}
\usepackage{booktabs}
\usepackage{multirow}

\usepackage[T1]{fontenc}
\usepackage[utf8]{inputenc}
\usepackage{newtxtext}
\usepackage{amsmath}
\usepackage{mathtools}
\usepackage{amsfonts}
\usepackage{bm}
\usepackage{amssymb}

\usepackage{newtxmath}

\usepackage[utf8]{inputenc}
\usepackage[T1]{fontenc}
\usepackage{xcolor}

\usepackage{tabularx}

\usepackage{bm}

\pdfoutput=1
\usepackage{color}
\definecolor{LinkColor}{rgb}{0.156,0.439,0.688}
\usepackage{hyperref}
\hypersetup{
colorlinks=true,
citecolor=LinkColor,
linkcolor=LinkColor,
urlcolor=LinkColor
}
\usepackage{multirow}

\newcommand{\typ}{\rm{typ}}
\newcommand{\toy}{\rm{toy}}

\usepackage{tikz}
\usepackage{textcomp}

\def\be{\begin{equation}}
\def\ee{\end{equation}}
\def\bea{\begin{eqnarray}}
\def\eea{\end{eqnarray}}

\begin{document}
\title{Extreme statistics as a probe of the superfluid to Bose-glass Berezinskii-Kosterlitz-Thouless transition}
\author{Jeanne Colbois}
\email{jeanne.colbois@cnrs.fr}
\affiliation{Institut Néel, CNRS \& Université Grenoble Alpes, 38000 Grenoble, France}
\affiliation{MajuLab, CNRS-UCA-SU-NUS-NTU International Joint Research Unit, Singapore}
\affiliation{Department of Materials Science and Engineering, National University of Singapore, 9 Engineering Drive 1, Singapore 117575}
\author{Natalia Chepiga}
\affiliation{Kavli Institute of Nanoscience, Delft University of Technology, Lorentzweg 1, 2628 CJ Delft, The Netherlands}
\author{Shaffique Adam}
\affiliation{Department of Physics, Washington University in St. Louis, St. Louis, Missouri 63130, USA}
\affiliation{Department of Materials Science and Engineering, National University of Singapore, 9 Engineering Drive 1, Singapore 117575}
\author{Gabriel Lemarié}
\email{gabriel.lemarie@cnrs.fr}
\affiliation{MajuLab, CNRS-UCA-SU-NUS-NTU International Joint Research Unit, Singapore}
\affiliation{Centre for Quantum Technologies, National University of Singapore, Singapore 117543, Singapore}
\affiliation{Department of Physics, National University of Singapore, Singapore 117542, Singapore}
\affiliation{Univ. Toulouse, CNRS, Laboratoire de Physique Th\'eorique, Toulouse, France}
\author{Nicolas Laflorencie}
\email{nicolas.laflorencie@cnrs.fr}
\affiliation{Univ. Toulouse, CNRS, Laboratoire de Physique Th\'eorique, Toulouse, France}
\date{\today}
\begin{abstract}
Recent studies of delocalization-localization transitions in disordered quantum chains have highlighted the role of rare, chain-breaking events that favor localization, in particular for high-energy eigenstates related to many-body localization. In this context, we revisit the random-field XXZ spin-1/2 chain at \emph{zero} temperature with ferromagnetic interactions, equivalent to interacting fermions or hard-core bosons in a random potential with attractive interactions. We argue that localization in this model can be characterized by chain-breaking events, which are probed by the extreme values of simple local observables, such as the on-site density or the local magnetization, that are readily accessible in both experiments and numerical simulations. Adopting a bosonic language, we study the disorder-induced Berezinskii-Kosterlitz-Thouless (BKT) quantum phase transition from superfluid (SF) to Bose glass (BG), and focus on the strong disorder regime where localization is driven by weak links.
Based on high-precision density matrix renormalization group simulations, we numerically show that extreme local densities accurately capture the BKT transition, even for relatively short chains ranging from a few dozen to a hundred sites. We also discuss the SF-BG transition in the weak disorder regime, where finite-size effects pose greater challenges. Overall, our work seeks to establish a solid foundation for using extreme statistics of local observables, such as density, to probe delocalization-localization transitions in disordered quantum chains, both in the ground state and at high energy.
\end{abstract}
\maketitle
\section{Introduction}

\subsection{Context}
Many-body quantum interacting systems in the presence of disorder are notoriously difficult to study~\cite{giamarchi_localization_1987,giamarchi_anderson_1988,doty_effects_1992,fisher_boson_1989,altshuler_quasiparticle_1997,jacquod_emergence_1997,gornyi_interacting_2005,basko_metalinsulator_2006,pietracaprina_shift-invert_2018}, yet they raise fundamental questions in quantum statistical physics and condensed matter. A famously challenging question is to characterize many-body localization (MBL)~(for recent reviews, see Refs.~\cite{abanin_colloquium_2019, alet_many-body_2018, nandkishore_many-body_2015,sierant_many-body_2025}), a possible alternative to thermalization.

In the infinite-size and infinite-time limit, the behavior of this potential phase of matter has sparked intense debate~\cite{doggen_many-body_2018,suntajs_quantum_2020,suntajs_ergodicity_2020,sels_dynamical_2021,sirker_particle_2022,weisse_operator_2025,sierant_thouless_2020,panda_can_2020,abanin_distinguishing_2021,sierant_challenges_2022,morningstar_avalanches_2022,berger_numerical_2024,colmenarez_ergodic_2024,colbois_statistics_2024,biroli_large-deviation_2024,scoquart_role_2024,laflorencie_cat_2025,scocco_thermalization_2024,szoldra_catching_2024,weisse_operator_2025,deroeck_absence_2025}, in particular in one dimension where quantum fluctuations play a major role.
In this context, the spin-1/2 XXZ chain in a random magnetic field is a central example:
\begin{equation}
H =  \sum_{i}  \left[\frac{1}{2}  (S_{i}^{+} S_{i+1}^{-} + S_{i}^{-} S_{i+1}^{+}) + \Delta S_{i}^{z} S_{i+1}^{z}\right] -\sum_i h_i S_{i}^{z},
\label{eq:ham}
\end{equation}
where $\Delta$ is the Ising anisotropy parameter and the $h_i \in [-W,W]$ are independent and uniformly distributed random variables. The above spin Hamiltonian Eq.~\eqref{eq:ham} is rather generic because it also describes interacting particles in a random potential, either spinless fermions~\cite{jordan_uber_1928} or hard-core bosons~\cite{matsubara_lattice_1956,girardeau_relationship_1960}. A useful point for later discussions is that the local magnetization in the spin chain Hamiltonian is directly related to the particle density at site $i$ as $S_i^z= n_i - 1/2$. 

The recent focus of the MBL field has largely been on the role of rare \emph{delocalizing} events such as avalanches seeded by ergodic regions~\cite{de_roeck_stability_2017,thiery_many-body_2018,morningstar_avalanches_2022,szoldra_catching_2024} and long-range many-body resonances~\cite{morningstar_avalanches_2022,biroli_large-deviation_2024,colbois_statistics_2024,laflorencie_cat_2025,giri_from_2025,scoquart_role_2024,scoquart_scaling_2025}. 
However, an opposite but relevant direction is to study the role of rare, highly \emph{localized} regions that slow down the dynamics and could even prevent transport close to the transition~\cite{luitz_extended_2016,luitz_ergodic_2017,luschen_observation_2017,gopalakrishnan_griffiths_2016,agarwal_rare-region_2017,pancotti_almost_2018,deroeck_absence_2025}, or break the chain~\cite{dupont_from_eigenstate_2019, laflorencie_chain_2020, samanta_tracking_2020, colbois_breaking_2023}.

Importantly, certain issues related to the MBL transition are connected to aspects of the ground-state phase diagram~\cite{dupont_many-body_2019,dupont_from_eigenstate_2019,takayoshi_dynamical_2022}. In particular, the Berezinskii-Kosterlitz-Thouless (BKT) universality class of the MBL transition has been theoretically discussed~\cite{dumitrescu_kosterlitz-thouless_2019,morningstar_renormalization-group_2019,goremykina_analytically_2019,suntajs_ergodicity_2020,morningstar_many-body_2020,niedda_renormalization-group_2024}, and numerically addressed using extreme magnetization statistics~\cite{laflorencie_chain_2020}, despite the relatively small system sizes accessible to mid-spectrum shift-invert diagonalization techniques~\cite{luitz_many-body_2015,pietracaprina_shift-invert_2018}. The key idea relies on a drastically different behavior of the distributions of local magnetizations in the ergodic regime --- where the variance shrinks, in agreement with the eigenstate thermalization hypothesis (ETH)~\cite{deutsch_quantum_1991,srednicki_chaos_1994,rigol_thermalization_2008};  and in the MBL regime --- where these distributions exhibit characteristic power-law tails~\cite{khemani_obtaining_2016,lim_many-body_2016,dupont_many-body_2019,dupont_from_eigenstate_2019,hopjan_detecting_2021,hopjan_many-body_2020,laflorencie_chain_2020,colbois_breaking_2023,colbois_interaction_2024}. 
In Ref.~\cite{laflorencie_chain_2020}, focusing on the extreme values of the local magnetization reveals a BKT-type scaling for a critical disorder slightly below the most recent extrapolated estimates~\cite{sierant_polynomially_2020,colbois_interaction_2024}.

In this context, it is natural to ask whether such ideas—reminiscent of entanglement bottlenecks~\cite{bauer_area_2013,luitz_long_2016} that are relevant to the ETH–MBL transition at high energy~\cite{laflorencie_chain_2020,dupont_from_eigenstate_2019,colbois_breaking_2023}—might also apply to ground-state physics. One aim is to validate the use of extreme statistics of local observables to detect and characterize BKT localization-delocalization transition on small system sizes: at high energy, numerics is limited to short chains (up to $L=22$). Does a similar analysis hold up to scrutiny for $L \gtrsim 100$ sites?

Specifically, we ask whether the extreme statistics of local densities are also relevant for the zero-temperature superfluid (SF) to Bose-glass (BG) localization transition.
Indeed, this SF-BG transition belongs to the BKT universality class~\cite{giamarchi_localization_1987,giamarchi_anderson_1988,refael_strong_2013}, and in its strong-disorder version~\cite{altman_phase_2004,altman_insulating_2008,refael_strong_2013,pielawa_numerical_2013} is known to be driven by effective \emph{weak links} that effectively cut the chain (see below for a definition). Thus, it provides a much more controlled setting in which to test, validate, and highlight the possible limitations of extreme statistics of local observables for studying delocalization-localization transitions. 

We therefore focus on the role of such rare localized events in the ground state of Eq.~\eqref{eq:ham}, with two objectives in mind: (1) exploring the relationship between extreme magnetization and weak links, and (2) show that extreme magnetization correctly captures the strong disorder transition.
\begin{figure}[t!]
    \includegraphics[width=\columnwidth]{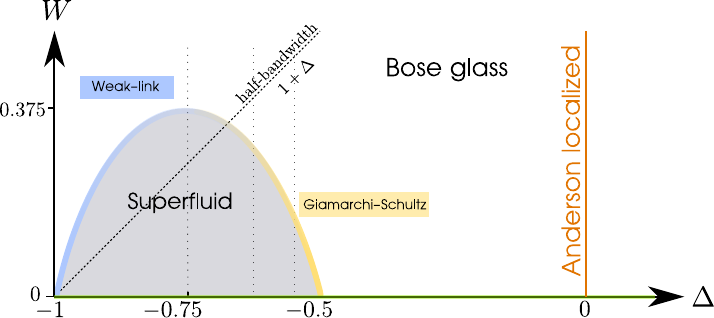}
    \caption{Sketch of the phase diagram of the spin-1/2 XXZ chain in a random field, Eq.~\eqref{eq:ham}, from Doggen {\it{et al.}}~\cite{doggen_weak_2017}. The green line indicates the Luttinger liquid regime ($-1 < \Delta < 1$) at $W=0$, which is a superfluid (SF) in the bosonic language. For $-1 < \Delta < -0.5$ the SF phase (in gray) remains stable against weak enough random field, while for $\Delta > -0.5$, any finite disorder immediately gives rise to a localized, gapless Bose glass (BG). The half-bandwidth $W^{\star} = 1+\Delta$ is expected to play a  qualitative role in controlling the nature of the BKT transition from the SF to the BG phase~\cite{doggen_weak_2017}. At strong disorder $W>W^\star$ the SF is destroyed by weak-links, while at weak disorder, it is destroyed by quantum phase slips (dubbed Giamarchi-Schulz~\cite{giamarchi_localization_1987}). In this work, we use extreme magnetization statistics to revisit the phase diagram along the vertical dotted lines.}
    \label{fig:PhaseDiag}
\end{figure}

\begin{figure*}
    \centering
    \includegraphics[width=\linewidth]{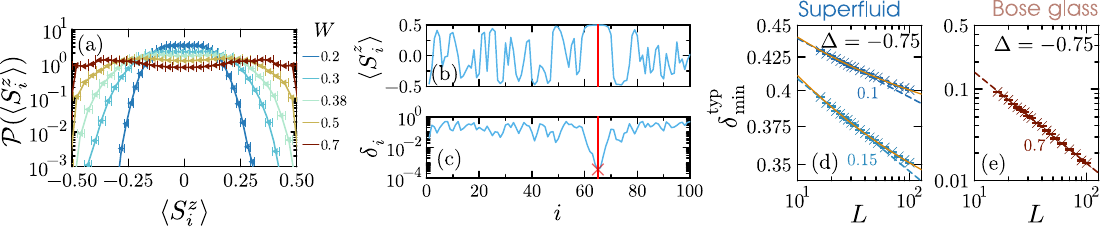}
    \caption{Extreme magnetization in the ground-state of the XXZ chain in random field at $\Delta = -0.75$. (a) Distribution of the expectation value of the local magnetization for $L = 100$ sites (5k independent samples). (b) Example of the expectation value of the local magnetization and (c) corresponding deviation for one sample, at $W=0.7$. (d) Decay of the typical minimal deviation in the SF phase ($W=0.1,\,0.15$), averaged over 10k samples for $L\leq 64$ and 5k samples otherwise. The dashed lines are fits to pure power laws, while the solid lines are fits to Eq.~\eqref{eq:deltafit} allowing a non-zero $\delta_{\infty}$. (e) The decay in the BG ($W=0.7$) is well captured by a pure power-law decay. Note that \emph{both} panels (d) and (e) are in log-log scale.}
    \label{fig:intro}
\end{figure*}

\subsection{Model, ground state, and weak links}
The random-field XXZ chain in Eq.~\eqref{eq:ham} has served as a test bed for studying disorder-induced ground-state localization transition in the presence of interactions~\cite{giamarchi_localization_1987,giamarchi_anderson_1988,doty_effects_1992}, thus going beyond the single-particle Anderson localization framework~\cite{anderson_absence_1958,evers_anderson_2008}.
This model is directly relevant for the physics of one dimensional arrays of Josephon Junctions~\cite{cedergren_insulating_2017,bard_superconductor-insulator_2017,kuzmin_quantum_2019,svetogorov_effect_2018,houzet_microwave_2019} and low-dimensional quantum magnets with impurities~\cite{hong_evidence_2010,ZHELUDEV2013740,dupont_numerical_2019}, and can be experimentally realized in ultracold atoms~\cite{derrico_observation_2014,gadway_glassy_2011}. More generally, it can be relevant to study the effects of disorder on superconductivity~\cite{chan_disorder_2025,charpentier_first-order_2025}. 

To set the problem, Fig.~\ref{fig:PhaseDiag} provides a schematic view of the ground-state phase diagram of Eq.~\eqref{eq:ham} as a function of disorder strength $W$, and interactions $\Delta$, focusing on the attractive regime ($-1<\Delta\le 0$)~\footnote{That is, ferromagnetic and easy-plane.}. 
In their seminal work, Giamarchi and Schulz \cite{giamarchi_localization_1987,giamarchi_anderson_1988} showed that weak disorder destabilizes the Luttinger liquid superfluid phase, leading to a Berezinskii–Kosterlitz–Thouless (BKT) transition into a Bose-glass insulator at the universal critical value $K_c=3/2$ of the Luttinger parameter. This result was further confirmed more recently by two-loop calculations~\cite{ristivojevic_phase_2012,ristivojevic_superfluid_2014}. As discussed in these works, this approach, being perturbative in the disorder, is well-controlled near the weak-disorder limit, but does not conclude on the universality of the transition at stronger disorder. 

A distinct strong-disorder regime, where different physical mechanisms dominate, has been proposed. Strong-disorder RG analyses \cite{altman_phase_2004,altman_insulating_2008,refael_strong_2013,pielawa_numerical_2013} highlighted the role of rare weak links, weak connections between superfluid puddles in the effective description of the chain. They proliferate and effectively cut the chain, yielding a non-universal critical value $K_c>3/2$. Building on this, Prokof’ev, Svistunov and collaborators \cite{pollet_classical-field_2013,pollet_asymptotically_2014,yao_superfluid-insulator_2016,pfeffer_strong_2019} incorporated the Kane–Fisher mechanism~\cite{kane_transmission_1992,kane_transport_1992}, whereby quantum fluctuations renormalize these weak links, rendering them either transparent or severed depending on the value of $K$. This leads again to a BKT transition, but governed by a non-universal critical value $K_c>3/2$.  

The existence of these two regimes has been confirmed by large-scale numerical simulations~\cite{hrahsheh_disordered_2012,doggen_weak_2017}. In particular, Doggen \emph{et al.} \cite{doggen_weak_2017} investigated the random-field XXZ chain using DMRG and QMC techniques, and reported a transition consistent with the Giamarchi–Schulz scenario with $K_c=3/2$ at weak disorder, while finding a distinct strong-disorder regime characterized by $K_c>3/2$. In this latter regime, power-law distributions were observed for several physical observables, supporting the relevance of the weak-link mechanism.

The qualitative picture distinguishing the weak-disorder (Giamarchi-Schulz) regime from the strong-disorder (weak-link)  behavior~\cite{doggen_weak_2017} is related to the half-bandwidth $W^{\star}(\Delta) = 1 + \Delta$, which corresponds to the saturation field in the clean XXZ chain. Indeed, in the clean case, the polarized ferromagnet obtained for $|h| > W^{\star}$ corresponds to an insulator in the bosonic language. Thus, in the presence of disorder, a weak link can take the form of a locally insulating region of length $r$ where all $h_{i} > W^{\star}$.  If $W > W^{\star}$, this occurs with probability $P(r) \sim (1- W^{\star}(\Delta)/W)^{r}$. In contrast, $W < W^{\star}$ prevents such an origin for weak links.

\subsection{Main results}
In this work, we mostly focus on the strong disorder regime of the SF-BG transition: we argue that the extreme magnetization probes a microscopic version of the weak link physics, and show that the local magnetization as a simple probe of localization correctly captures both the location and scaling of the transition, even with modest system sizes in the range $L=14-100$~\footnote{Both the memory and time complexity of DMRG scale linearly with the system size at \emph{fixed} bond dimension. However, with larger system sizes, larger bond dimensions may be needed, and stronger entanglement bottlenecks arise, which implies that the convergence requires more sweeps, thereby increasing the numerical cost even further.}, compared to $L = 300 - 512$ sites required in Ref.~\cite{doggen_weak_2017}. We also discuss finite-size effects at weaker interactions and characterize the behavior of this observable in the weak disorder (Giamarchi-Schulz) regime of the transition.

The importance of these results is threefold.  First, they put on a stronger footing an approach previously applied to the ETH-MBL transition with available sizes $L \leq 22$~\cite{laflorencie_chain_2020}. Second, they highlight the contrast between the weak-disorder and  the strong-disorder regimes of the transition, and might revive the possibility of two different Bose glass phases~\cite{giamarchi_localization_1987,ristivojevic_superfluid_2014,doggen_weak_2017}. Last but not least, they provide a basis for experimentally studying the role of entanglement bottlenecks through much simpler observables such as the local magnetization or density.

The key idea is summarized in Fig.~\ref{fig:intro}. For each disorder realization of the XXZ chain, we focus on the local magnetization $\langle S_i^{z} \rangle =  \langle n_i \rangle - 1/2 $ in the ground state. The distributions of the magnetization show a clear evolution with increasing disorder strength (Fig.~\ref{fig:intro}a), with a higher probability of finding very strongly polarized sites at strong disorder. Quantum fluctuations however forbid a perfectly polarized site for any finite size ($|\langle S_i^{z} \rangle| < 1/2$)~\footnote{That is, in the particle language a perfectly empty or occupied site is forbidden.}. To quantitatively characterize the differences between the distributions of the magnetization in the SF and the BG phase, it is useful to focus on the most polarized site in the chain (Fig.~\ref{fig:intro}b), i.e. the one where the deviation to perfect polarization
\begin{equation}
\label{eq:deltadef}
    \delta_{i} = \frac{1}{2}- |\langle S_i^{z} \rangle| = \frac{1}{2}- |\langle n_i \rangle -1/2|
\end{equation}
is minimal (Fig.~\ref{fig:intro}c). For reasons that will become clear in Sec.~\ref{sec:noninteracting}, we focus on the typical value of the minimal deviation
\begin{equation}
\delta^{\typ}_{\min} = \exp{ \overline{\ln (\min_{i} \delta_{i}})},
\end{equation}
where the bar indicates disorder average.
Studying its scaling for increasing system sizes $L$ reveals a fundamental difference between the SF and the BG phases in the strong disorder regime. Indeed, in the SF, we find that the minimal deviation decays to a non-zero value in the thermodynamic limit (Fig.~\ref{fig:intro}d, nuanced below), while at strong disorder in the BG, it goes to zero as a power law (Fig.~\ref{fig:intro}e). Both behaviors are captured by the following phenomenological description of our finite length data
\begin{equation}
\label{eq:deltafit}
    \delta_{\min}^{\typ}  = BL^{-\tilde{\gamma}} + \delta_{\infty},
\end{equation}
with $\delta_{\infty}$ going to zero in the BG. Even for modest system sizes, we show that these contrasting behaviors  reliably capture the SF-BG transition when it is driven by weak links. 

The rest of the paper is organized as follows. In Section~\ref{sec:noninteracting} we focus on the non-interacting $\Delta = 0$ limit (many-body Anderson case). We recall essential points at high energy and summarize key new results in the ground state, on the basis of exact diagonalization on large chains with periodc boundary conditions. The remainder of the paper focuses on finite attractive interactions, $\Delta < 0$, for which the results are obtained using DMRG on chains with open boundary conditions. In Section~\ref{sec:chainbreaking}, we present two ways to use extreme magnetization to spot the transition in the interacting model in the strong disorder (weak-link) regime, specifically at $\Delta = -0.75$.  Section~\ref{sec:BKT} provides the BKT scaling analysis of the transition based on this probe. In Section~\ref{sec:discussion}, we discuss a possible interpretation of our results, as well as a conjecture for the Giamarchi-Schulz (weak-disorder) regime, where the extreme statistics behave very differently. A summary and outlook are given in Section~\ref{sec:outlook}. The main appendix, App.~\ref{app:AL}, provides details related to Section~\ref{sec:noninteracting}, and  significantly extends the discussion of the new results for the $\Delta = 0$ ground state. Other appendices clarify the DMRG parameters, the number of samples, and the minimal deviation selection to avoid edge effects (App.~\ref{app:DMRG}), details on the scaling procedure (App.~\ref{app:scaling}), and finally the main characteristics of the full magnetization distributions (App.~\ref{app:distributions}). 

\section{Spin freezing and chain breaking in the non-interacting case}
\label{sec:noninteracting}
We first focus on the Hamiltonian Eq.~\eqref{eq:ham} in the $\Delta = 0$ limit. There, the model is an XX chain in random field~\footnote{The first term is $\sum_{i} (S_i^{x}S_{i+1}^{x} + S_i^{y}S_{i+1}^{y})$}, which maps to an Anderson localized chain of non-interacting spinless fermions in a random potential through the Jordan-Wigner transformation~\cite{anderson_absence_1958, Mott_1961,jordan_uber_1928}. The local magnetization is related to the local fermion density, the total magnetization conservation is equivalent to the particle number conservation, and the $S^{z}_{\rm tot} = 0$ sector corresponds to the Anderson localized chain will half the levels filled. In the discussion below, we use the term ``many-body'' Anderson localization despite the non-interacting spinless fermion model to highlight the role of the filling. Section~\ref{sec:spinfreezing} briefly recalls key elements linking extreme magnetization, chain breaking and localization length at high energy, previously discussed in Refs.~\cite{dupont_many-body_2019,dupont_from_eigenstate_2019,laflorencie_chain_2020,colbois_breaking_2023}. We then provide in Section~\ref{sec:groundstate} central results in the ground state of this non-interacting case.

\subsection{Known results at high energy}
\label{sec:spinfreezing}
For high-energy eigenstates in the non-interacting, half-filled Anderson localized case, the typical minimal deviation Eq.~\eqref{eq:deltadef} decays as a power-law of the system size $L$ for any finite disorder $W$~\cite{colbois_breaking_2023}:
\begin{equation}
    \delta^{\typ}_{\min}(L) = A(W) L^{-\gamma(W)}.
    \label{eq:power-law}
\end{equation}
This implies that in the thermodynamic limit, there is at least one spin in the chain that is perfectly polarized. 

This behavior is well captured by a simple toy model~\cite{dupont_from_eigenstate_2019, laflorencie_chain_2020, colbois_breaking_2023}, which we briefly recall here.

Building on the spin-particle correspondence, we model the random-field XX chain (which corresponds to a many-body Anderson localized chain) by a collection of half-filled localized orbitals (labeled by $m$), each centered on a different site $i_m$ and all with the \emph{same} localization length $\xi_{\rm{AL}}$: $|\phi_m(i)|^2 = A_m \exp\left(- \frac{|i - i_m|}{\xi_{\rm{AL}}}\right)$. This is a minimal description of the many-body Anderson insulator, which nevertheless captures the essential features of extreme statistics of local densities. In this  simplified framework, the most polarized site is located in the middle of the longest series {$\ell_{\max}\sim \ln L$} of neighboring orbitals which are all either empty or all occupied. It is easy to show that the power-law exponent is directly related to the { localization} length {$\xi_{\rm{AL}}$} (see Appendix~\ref{app:AL} and Refs.~\cite{colbois_breaking_2023,dupont_from_eigenstate_2019} for details):
\be
\delta_{\min}^{\toy} \propto e^{-\frac{\ln L}{2\xi_{\rm{AL}} \ln 2}} =  L^{-\frac{1}{2\xi_{\rm{AL}} \ln 2}}.
\label{eq:toy}
\ee
Upon lowering the disorder, the toy model breaks down for $W \lesssim 1.6$ ($\xi_{\rm AL} \gtrsim 1$), but the power-law decay remains valid with an exponent $\gamma > \frac{1}{2 \xi \ln 2}$ for large enough system sizes.

\subsection{Ground state properties}
\label{sec:groundstate}
We now summarize a few main messages on the minimal deviation from perfect polarization in the ground state of the XX chain. These key results support the discussion of the Giamarchi-Schulz regime in Section~\ref{sec:discussion}. We refer the reader to Appendix~\ref{app:AL} for a detailed presentation of our results at $\Delta = 0$, in particular for the properties of the distributions. 

The main difference between the high-energy eigenstates of the XX chain in the $S^z_{\rm tot} = 0$ sector and its ground state is which single-particle Anderson-localized orbitals are occupied. In the latter, the lower half of the spectrum is full, and the full range of the localization lengths in the single-particle spectrum appear. Indeed, both states with very short length $\xi_{\min}$ at the band edge, and states with { much larger} localization length $\xi_{\max}$ in the middle of the spectrum are occupied. For strong disorders, when all localization lengths are short, we do not expect fundamental differences in the decay of the typical minimal deviation, and it is indeed captured by Eq.~\eqref{eq:power-law}. The main difference is that the minimal deviation is controlled mostly by much shorter localization lengths than the average $\xi_{\mathrm{AL}}$ (see Appendix~\ref{app:AL} for a detailed argument). As a result, we observe that the exponent $\gamma_{\rm GS}$ controlling the decay in the ground state is larger than $\gamma$ controlling that at high energy, for the same disorder. This result is valid for $W \gtrsim 1.2$.

For weaker disorder, strong finite-size effects appear. From the toy model, it is naturally expected that the power-law decay of the minimal deviation cannot be captured correctly when the localization length far exceeds {$\ell_{\rm{max}}\sim\ln L$}, and this is indeed what we observe. In practice, this suggests a lower-bound on the exponent $\gamma$ that we can capture correctly { for a given system size $L$}: $\gamma \gtrsim 0.1$ for the large ({ $L=4000$} sites) PBC chains we study with exact diagonalization in Appendix~\ref{app:AL}, and $\gamma \gtrsim 0.2$ for the more modest sizes { ($L\approx 100$)} we reach in the interacting systems in the rest of the paper.

\section{Extreme statistics and strong disorder transition}
\label{sec:chainbreaking}

We now focus on the strong disorder regime of the SF-BG transition. We present two procedures to spot the SF-BG transition by taking advantage of the two contrasting behaviors seen in Fig.~\ref{fig:intro}. The first one (Section~\ref{sec:effective}) is inspired by the approach used for MBL in Ref.~\cite{laflorencie_chain_2020}, which relies on the crossing of size-dependent effective exponents. The second one comes from the observation that the numerical results are well described by Eq.~\eqref{eq:deltafit} with $\delta_\infty > 0$ in the SF, see Section~\ref{sec:limitingvalue}.

\begin{figure*}
    \centering
    \includegraphics[width=\linewidth]{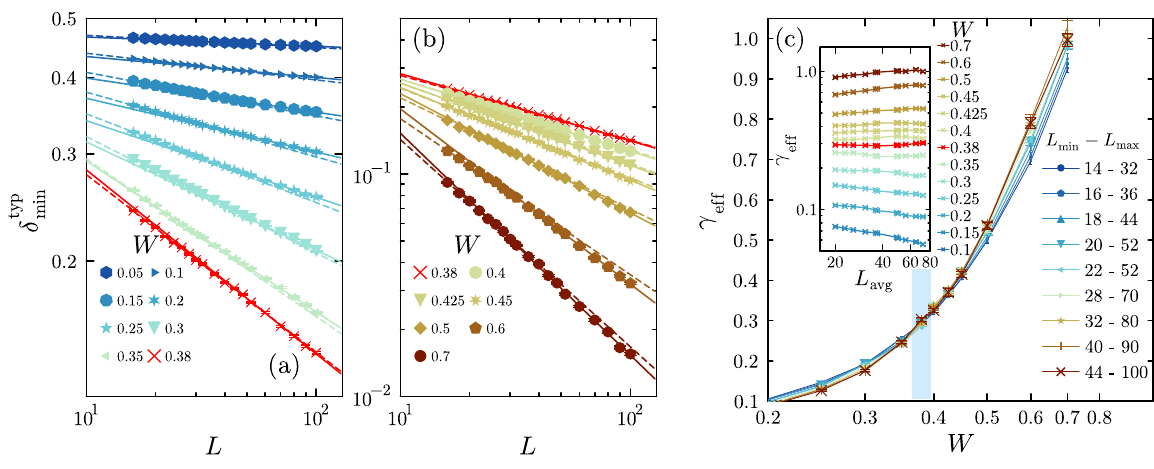}
    \caption{Decay of the typical minimal deviation at $\Delta = -0.75$ (weak-link/strong disorder transition) and effective exponent.  The critical disorder strength is $W = 0.375 \pm 0.015$~\cite{doggen_weak_2017}. (a) and (b) Decay of the typical minimal deviation, in log-log scale. Lines are fits for $\ln \delta_{\min}^{\typ} = \ln A - \gamma_{\rm eff} \ln L$ for $L \in [14,28]$ (dashed) and $L \in [48,100]$ (full). Below the critical disorder strength (a), the decay is slower than a pure power-law decay (in agreement with $\gamma_{\rm eff} \rightarrow 0$, see panel (c)), whereas at and above the critical disorder strength (b), $\delta_{\min}^{\typ}(L)$ decays as (or slightly faster than) a pure power-law. (c) Effective power-law exponents $\gamma_{\rm eff}$ from sliding fits, log-linear scale. The sizes for sliding fits are chosen such that $\ln L_{\max} - \ln L_{\min}$ is roughly constant, and the fits for the smallest sizes and largest sizes correspond to those shown in the first panels. Inset: same data, shown for various disorders as a function of the mid-point of the fitting interval and highlighting the trend inversion between increasing and decreasing $\gamma_{\rm eff}$, and the saturation for $\gamma_{\rm eff}$ for large system sizes above the transition. }
    \label{fig:Delta-075}
\end{figure*}

\subsection{Crossing of the effective exponents}
\label{sec:effective}
To  locate the transition in the strong-disorder (weak-link) regime, we use the \emph{effective} exponent $\gamma_{\rm eff}(L)$ extracted from successive power-law fits using sliding fitting windows involving larger and larger system sizes. The idea is that for a pure power-law decay, $\gamma_{\rm eff} = \gamma $ is independent of the fitting window. In contrast, for a slower-than-power-law decay, $\gamma_{\rm eff} (L)\rightarrow 0$ with increasing sizes, and for a faster-than-power-law decay, $\gamma_{\rm eff} (L) $ grows. 

As already seen in Fig.~\ref{fig:intro}(e), we can reasonably expect the minimal deviation in the Bose glass to behave similarly as in the Anderson localized case ($\Delta =0$). Namely, we expect it to decay as a power-law with increasing system size ($\delta_{\infty} = 0$ in Eq.~\eqref{eq:deltafit}), with a well-defined exponent $\gamma$. However, similar analysis at high energy (both in the non-interacting model~\cite{colbois_breaking_2023} and in the random-field Heisenberg chain~\cite{laflorencie_chain_2020}), and in the non-interacting ground state (Appendix~\ref{app:AL}) show that $\gamma$ is not fully free of finite-size effects \emph{even} on the localized side. Rather, the effective exponent $\gamma_{\mathrm{eff}}(L)$ grows to $\gamma$ from below~\footnote{We interpret this as follows, in the non-interacting case. There, $\gamma$ corresponds to an inverse localization length. For small samples,  $\xi$ is naturally expected to crossover towards smaller values when $L$ grows, yielding an opposite finite-size effect for $\gamma_{\rm eff}(L)$, which grows slightly with $L$.}. In contrast, in the SF, we expect the minimal deviation to go to a non-zero value and $\gamma_{\rm eff}(L) \rightarrow 0$ as $L\rightarrow \infty$. The point where the behavior of $\gamma_{\rm eff}(L)$ changes is therefore expected to correspond to the transition. 

To validate this approach, we study this behavior at $\Delta= -0.75$ in Fig.~\ref{fig:Delta-075}. Panel~(a) shows the typical minimal deviation as a function of system sizes in the SF. The dashed lines correspond to fits on the smaller system sizes and the solid lines to fits on the larger system sizes (c.f. caption), highlighting the finite-size crossover of the effective exponent. Panel (b) presents the decay of the minimal deviation in the BG. The log-log scale used in both panels highlights the power-law behavior in the localized phase, contrasting with the SF. For $W \geq 0.5$,  a slightly slower decay of $\delta^{\typ}_{\min}(L)$ for smaller system sizes is visible, in agreement with the expected finite-size effect $\gamma_{\mathrm{eff}}\rightarrow \gamma$ from below. 

As a result of these contrasting behaviors in the SF and the BG, Fig.~\ref{fig:Delta-075}c shows a crossing of $\gamma_{\rm eff}(L)$ in the grey region, which nicely coincides with the SF-BG transition point $W_c=0.375(15)$ previously extracted using a standard BKT analysis of the superfluid stiffness obtained from quantum Monte Carlo simulations~\cite{doggen_weak_2017}. 

At very strong disorder, we also find the expected logarithmic dependence $\gamma_{\mathrm{eff}} \propto \ln(W/W_{\star})$ (see Appendix~\ref{app:distributions}). Indeed, at high energy, and if the disorder is strong enough, one can argue using simple perturbative ideas~\cite{dupont_from_eigenstate_2019} that the freezing exponent $\gamma(W)\sim \ln W$, which matches the behavior of the localization length at strong disorder. 

{ Finally,} the fact that $\gamma$ is finite at the transition tells us that this freezing exponent \emph{cannot} be simply interpreted as an inverse localization length in the interacting case - or, at least, not at criticality. Indeed, both the weak-link and the GS critical points are expected to show a finite superfluid density at the transition~\cite{doggen_weak_2017} and a diverging correlation length~\cite{giamarchi_anderson_1988,ristivojevic_phase_2012,ristivojevic_superfluid_2014}. The result that $\gamma_{\infty} > 0$ at $W_c$ thus seems surprising at first glance. We clarify in Section~\ref{sec:discussion} that this result is compatible with algebraically decaying long-range correlations provided that there is a healing at long distance across the highly polarized site exactly at the transition.

\subsection{Large system limit of the minimal deviation}
\label{sec:limitingvalue}
\begin{figure}
    \centering
    \includegraphics{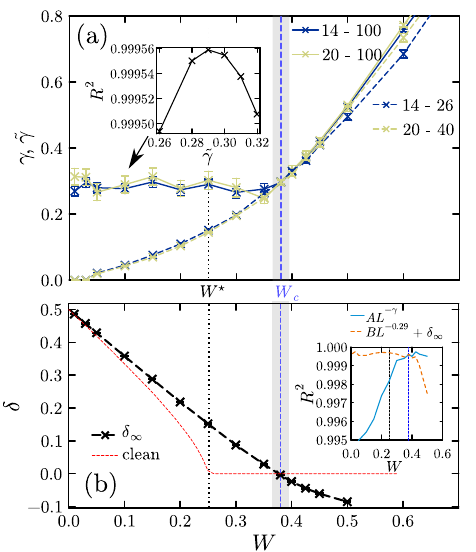}
    \caption{$\Delta = -0.75$: Comparing pure power-law fits with exponent $\gamma$ to fits with an additional constant and exponent $\tilde{\gamma}$. (a)  Exponents comparison. $\tilde{\gamma}$ is obtained from a 3-parameter fits, hence the larger errors, but all sizes from $L_{\min}\in {14, 20}$ to $L_{\mathrm{max}} = 100$ are included in the fit. $\gamma$ is obtained as in Fig.~\ref{fig:Delta-075}. Inset: optimal value for $\tilde{\gamma}$ at $W = 0.1$. (b) Constant $\delta_{\infty}$ from the fit Eq.~\eqref{eq:deltafit} (crosses) for a fixed exponent $\tilde{\gamma} = 0.29$, compared to the clean value for $\delta = 1/2-m$ (red dashed line). Inset: comparison of fit qualities between this and a pure power-law fit.}
    \label{fig:extrapolated}
\end{figure}

As sketched in the introduction, an alternative way to spot the transition is to compare pure power-law fits to fits of the form Eq.~\eqref{eq:deltafit}, i.e. with an additional constant. As shown in Fig.~\ref{fig:intro}, the latter seem to capture well the results in the SF phase.  We now make this more quantitative. In this type of fit, the exponent $\tilde{\gamma}$ of the power-law part is \emph{not} expected to go to zero. Instead, Fig.~\ref{fig:extrapolated}(a) suggests that $\tilde{\gamma}$ is finite and depends only weakly on $W$ in the SF phase for $\Delta = -0.75$. This motivates the following analysis: 

(i) at weak disorder (i.e. deep in the SF phase), we find the exponent providing the best fit for the power-law part of the decay, see inset Fig.~\ref{fig:extrapolated}(a).

(ii) we then fix this exponent to $\tilde{\gamma}= 0.29$ so that we can compare two different 2-parameter fits : a pure power-law form Eq.~\eqref{eq:power-law} {\it{vs.}} with flexible exponent $\gamma$, and a power-law with a \emph{fixed} exponent $\tilde{\gamma}_{\rm const}= 0.29$ plus an additional constant $\delta_{\infty}$, Eq.~\eqref{eq:deltafit}.

The result for $\delta_{\infty}$ with this fixed exponent is shown in Fig.~\ref{fig:extrapolated}(b). For $W \leq W_c$, we expect $\tilde{\gamma}_{\rm const} = 0.29$ to be close to $\tilde{\gamma}$ (see Fig.~\ref{fig:extrapolated}(a)) and the resulting $\delta_{\infty}$ to be a good estimate of the infinite-size limit of the minimal deviation. We first remarkably observe that $\delta_{\infty}$ passes exactly through zero at the transition point $W_c \approx 0.38$. For $W > W_c$, $\tilde{\gamma}_{\rm const} = 0.29$ is no longer a good choice for the exponent, and accordingly, the negative $\delta_\infty$ indicates that the fit is poor. An alternative fit (not shown) in which $\tilde{\gamma}$ is \emph{not} fixed results in $\delta_{\infty}=0$ above the transition (but much larger errors on $\delta_{\infty}$ for $W \leq W_c$). We confirm this by comparing the fixed power fit to a pure power-law Eq.~\eqref{eq:power-law} with flexible exponent. The former provides a consistently better description than the pure power-law all the way up to $W = W_c$, as shown by the larger $R^2$ values in the inset~\footnote{The $R^2$ is the coefficient of determination. It is obtained by comparing the residuals of the data $y$ with respect to the fit $f$ from those with respect to the average : 
\begin{equation}
    \label{eq:Rsquared}
    R^2 := 1 - \frac{\sum_i {(y_i -f_i)^2}}{\sum_{i}(y_i -\overline{y})^2}.
\end{equation}
$R^2 = 1$ ($R^2 \ll 1$) indicates a perfect (bad) fit.}. In contrast, for $W > W_c$, the pure power-law with flexible exponent offers a better fit, consistent with the conclusion that $\delta_{\infty} = 0$ in this regime~\footnote{We checked that this is also the case at stronger interactions ($\Delta = -0.875$). We note (data not shown) that the very small value of $\delta_{\infty}$ close to the transition  for $W^{\star}(\Delta) < W <W_c(\Delta)$ can render the analysis difficult. Correct conclusions rely on having (1) a sufficiently high number of samples ($\gtrsim1500$) and (2) a sufficiently large set of different sizes to perform reliable fits.}. Thus, the behavior of $\delta_{\infty}$ in Eq.~\eqref{eq:deltafit} sharply locates the strong-disorder transition. Note that the result that $\delta_{\infty} > 0$ below the transition also implies that $\lim_{L \rightarrow \infty} \gamma_{\rm eff}(L) = 0$, while $\delta_{\infty} = 0 $ for $W \geq W_c$ is compatible with $\lim_{L \rightarrow \infty} \gamma_{\rm eff}(L) = \gamma > 0$ at and above the transition (see inset of Fig.~\ref{fig:Delta-075}(c)), suggesting a jump of the (pure) power-law exponent of Eq.~\eqref{eq:power-law} at the transition in the infinite size limit.

These results raise three main questions. First, beyond locating the transition, can the extreme magnetization also capture the nature of the transition, in particular, the BKT scaling expected in the present case? Second, can one have a consistent interpretation  of the value of $\delta_{\infty}$ at the transition (and thus of the behavior of the power-law exponent $\gamma$)? And third, would such an approach be as successful in the weak-disorder Giamarchi-Schulz regime? We address the first question in Sec.~\ref{sec:BKT}, and discuss the two others in Sec.~\ref{sec:discussion}.

\section{Scaling in the strong disorder regime}
\label{sec:BKT}

\begin{figure*}
    \centering
    \includegraphics[width=0.9\textwidth]{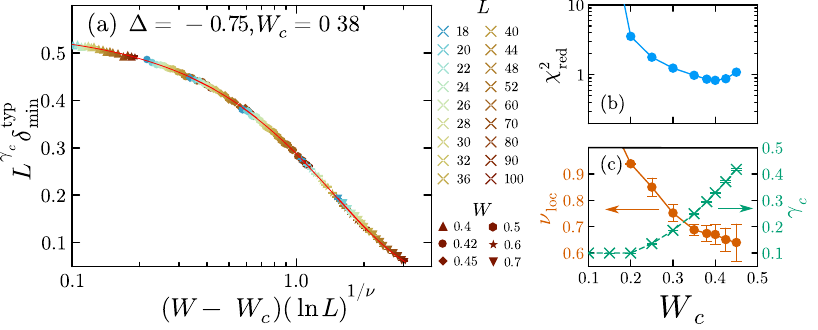}
    \caption{\label{fig:Delta-075collapse}
    Finite-size scaling analysis of the typical minimal deviation in the strong disorder regime of the transition, on the BG side. The scaling analysis is performed for various candidate values for the critical disorder $W_c$. (a) Scaling collapse for fixed $W_c = 0.38$. Different symbol shapes correspond to different disorder strengths, and different colors to different system sizes. For this value of $W_c$, corresponding to the critical disorder discussed in the previous section, the best collapse is obtained with $\gamma(W_c) = \gamma_c = 0.294 \pm 0.002$ and $\nu_{\rm loc} = 0.67 \pm 0.03$ (errors are estimated from a parametric bootstrap). The red line is the scaling function Eq.~\eqref{eq:scalingfit} with fitted parameters.  (b) Reduced chi-square for the best fit as a function of the chosen critical disorder strength $W_c$.  A very large $\chi^2_{\rm red}$ indicates a poor fit, while one of order 1 indicates a reasonable fit: for $W_c \in [0.35, 0.45]$ and in particular for $W_c = 0.38$, the fit is good. (c) Exponent $\nu_{\rm loc}(W_c)$ and exponent $\gamma_c(W_c)$ as a function of the same critical disorder strength.}
\end{figure*}

We focus on the BG side of the transition in the strong disorder regime. The power-law behavior of the typical minimal deviation demonstrated in the previous section (Fig.~\ref{fig:Delta-075}) suggests a logarithmic scaling of the form $\frac{\ln L}{\lambda}$. Instances of such scaling forms (but with either the volume of the Hilbert space or the volume of the random graph playing the role of $L$), can be found in Refs.~\cite{garcia-mata_two_2020,garcia-mata_scaling_2017,garcia-mata_critical_2022,mace_multifractal_2019,tikhonov_statistics_2019}. An analogous form in time was recently discussed in small-world graphs~\cite{chen_critical_2025}. More directly relevant here, a logarithmic scaling for the extreme magnetization on the localized side of the ergodic-MBL transition was discussed in Ref.~\cite{laflorencie_chain_2020}.
In that work as in the present case, it is justified by the expected behavior
\be
  \frac{\delta^{\typ}_{\min}(W)}{\delta^{\typ}_{\min}(W_c)} = \frac{A(W)}{A(W_c)}e^{-\left(\gamma(W_c)-\gamma(W)\right) \ln L}. 
\ee
Thus, ignoring the disorder-dependency of the amplitude $A(W)$ for the moment, we can write a scaling hypothesis as
\be
  \delta^{\typ}_{\min} = \delta^{\typ,c}_{\min} f\left(\frac{\ln L}{\lambda} \right)
  \label{eq:scaling1}
\ee
with $f(X) = D e^X $  for large $\ln L \gg \lambda$. This, together with the interpretation of $\gamma$ as an inverse localization length in the non-interacting case, suggests that $\ln(L)$ here plays a role similar to that of a length scale in usual collapse analysis. 

To validate the above hypothesis, we use a controlled finite-size scaling analysis~\cite{slevin_corrections_1999,rodriguez_multifractal_2011,laflorencie_chain_2020}. The idea is to seek a global fit of the following form for all the data for $W > W_c$ for all sizes:
\be
L^{\gamma_c} \delta^{\typ}_{\min}  = F\left( (\ln L)^{\frac{1}{\nu_{\rm loc}}} \rho(W-W_c)\right),
\label{eq:scaling2}
\ee
where  $\rho(W-W_c)  \sim (W-W_c)$ close to the critical point. This is done through a systematic fit in the form of a Talylor expansion (see Appendix~\ref{app:scaling} for details). For fixed critical disorder $W=W_c$, a collapse with such a scaling gives a divergence 
\be
    \lambda \sim (W-W_c)^{-\nu_{\rm loc}}.
\ee

We systematically perform this analysis for various tentative critical disorders $W_c$, as shown in Fig.~\ref{fig:Delta-075collapse}. 
In panel (a), the result for $W_c = 0.38$ is shown with the associated scaling function $F$. The reduced $\chi^{2}$ values as a function of $W_c$ are shown in panel (b),  illustrating that the fits are of good quality for $W_c \in [0.35, 0.45]$. 
For each of these fits, we show the resulting exponent $\nu_{\rm loc}$ controlling the power-law divergence of the logarithmic length scale $\lambda$ near the critical point (panel c). However, because of the scaling form Eq.~\eqref{eq:scaling1}, this power-law divergence with $(W-W_c)$ corresponds to an effectively exponentially diverging length scale controlling the finite-size effects. This is in agreement with the expected BKT scenario, even though the obtained exponent $\nu_{\rm loc} =  0.67 \pm 0.03$ is larger than the usually expected $\nu_{\rm loc}= 1/2$. It is often challenging to obtain such exponents precisely, and we do not a priori believe this difference to be very meaningful. This result also shines light on the similar analysis for the high-energy MBL transition in the random-field Heisenberg chain model mentioned in the introduction~\cite{laflorencie_chain_2020}. Indeed, while this analysis gives $\nu_{\rm loc} \sim 0.52$ for a critical disorder strength $h_c= 4.2(5)$, the results are also perfectly compatible with $\nu_{\rm loc}\sim 0.6$ for the stronger critical disorder strength $h_c\approx 5$ recently discussed in the MBL literature on this model and extracted from similar numerics~\cite{doggen_many-body_2018, sierant_polynomially_2020, colbois_interaction_2024}. 
We note that the critical value $\gamma_c$ (also shown in panel (c)) depends on $W_c$ and matches the $\gamma(W)$ obtained in Fig.~\ref{fig:Delta-075} at $W=W_c = 0.38$, as should be expected.

Note that $A(W)$ actually depends on disorder, with  $\ln \frac{A(W)}{A(W_c)}$ playing a role in the scaling. In fact, in the crude approximation where the full distribution of deviations is modeled as a pure power-law, $A(W)$ is related to $\gamma(W)$ through the normalization of the distribution. As we discuss further in Appendix~\ref{app:scaling}, this type of term, also present in the non-interacting case~\cite{colbois_breaking_2023}, can be tentatively characterized by irrelevant logarithmic corrections of the form $f(\ln L/ \lambda) \rightarrow f(\ln L/\lambda) + (\ln L)^{-1} h(\ln L/\lambda)$~\cite{laflorencie_chain_2020}. This does not change the results we obtain for the BKT-like scaling exponent $\nu_{\rm loc}$.

\section{Discussion}
\label{sec:discussion}

\subsection{Extreme magnetization: a local probe of weak-link physics?}

Given the successful analysis of the strong-disorder (weak-link-driven) transition using minimal deviations, it is tempting to ask whether one can interpret spin freezing and chain breaking as a local probe of weak-link physics. To this end, we first recall a few key points about the weak links in the existing literature.

In the strong disorder scenario \emph{\`a la} Altman {\it{et al.}}~\cite{altman_phase_2004,altman_insulating_2008,altman_superfluid-insulator_2010,refael_strong_2013}, the weak links are weak Josephson couplings in the effective description of the model reached after several strong-disorder-RG steps. The exponent $\alpha$ characterizing the power-law tails of weak-links $P(J) \sim J^{\alpha}$ is therefore a size-dependent quantity, and the critical point is characterized by a universal value of $\alpha$ (but a non-universal value of $K_c$). Taking into account a Kane-Fisher healing of the weak-links, the scratched-XY universality scenario~\cite{pollet_classical-field_2013,pollet_asymptotically_2014,yao_superfluid-insulator_2016,lemarie_kane-fisher_2019,pfeffer_strong_2019} also suggests that $K_c$ is non-universal, but  depends on $\alpha$ as~\cite{pfeffer_strong_2019} $K_c = \frac{\alpha+1}{\alpha}$\footnote{This expression is  different from the result stated in the appendix of Ref.~\cite{doggen_weak_2017}, but does not change the conclusion that this dependence is not in agreement with the power-law tail of the correlators at the transition.}. 

Crucially, in the numerical literature, the weak links are not probed \emph{directly}. Rather, their exponent is probed indirectly through the effect of the weakest link on a macroscopic observable. For instance, QMC is used to probe the weak-link exponent by studying the power-law tail of the superfluid stiffness distributions either directly~\cite{pielawa_numerical_2013,doggen_weak_2017} or by looking at rare events~\cite{yao_superfluid-insulator_2016}. Ref.~\cite{doggen_weak_2017} also focuses on the tail of the distribution of the one-body density matrix (transverse correlators in spin language).  

In these analyses, an underlying hypothesis is that the weakest link controls the tail of the distributions at long distances. While this makes sense in the SF, the proliferation of weak-links in the Bose glass naturally gives rise to a breakdown of the power-law tails, both in the superfluid stiffness and in the long-range correlator~\cite{doggen_weak_2017}. 

Instead, as we have seen, the extreme magnetization captures a pure power-law decay on the Bose-glass side of the transition (at least at strong enough disorder, see below), converging to a well-defined exponent for sufficiently large sizes. Furthermore, the extreme magnetization events also seem to exhibit healing in the superfluid phase. It may therefore be tempting to interpret the extreme magnetization events, related to entanglement bottlenecks, as a local probe of the weak link physics.

However, a proper identification between microscopic entanglement bottlenecks and weak-links would require an analysis of the impact of such bottlenecks for the emergence of \emph{effective} weak couplings between superfluid puddles in the RG procedure.  A key point is what happens at the strong-disorder transition. As we observed at $\Delta = -0.75$, $\delta_{\infty} \rightarrow 0$ and $\gamma > 0$ at the transition. Yet, it is known that the critical point is superfluid, so the correlations at long distances must be (and are) healed: at this point, the correlations across the weakest link are affected only at short distances by its presence.

This suggests a crucial difference between the microscopic observable used here, which goes to zero at the transition, and the renormalized weak links, which have to be healed at the transition for the critical point to remain superfluid.

\subsection{Giamarchi-Schulz regime}

\begin{figure}[b!]
    \includegraphics[width=\columnwidth]{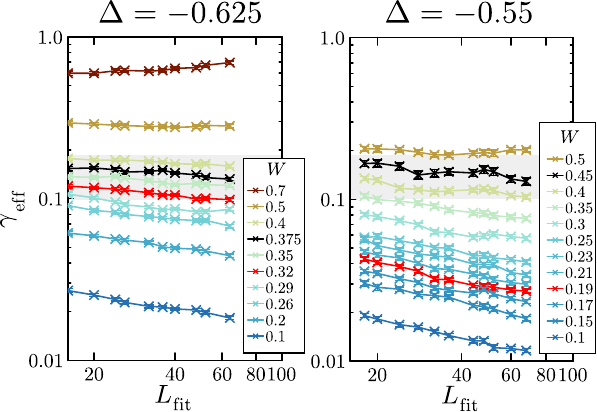}
    \caption{\label{fig:gammaeff} Effective exponent $\gamma_{\rm eff}$ in the Giamarchi-Schulz regime of the transition. The gray region indicates the order of magnitude of $\gamma_{\rm eff}$ below which finite-size effects are expected to be too significant for our analysis. The red (black) curves are the data at $W^{\star}$ ($W_c$), respectively. Here, $L_{\mathrm{fit}}$ is the middle point for the sizes taken into account in the sliding fit.}
    \label{fig:fitgammaeff}
\end{figure}

\begin{figure}[t!]
    \centering
    \includegraphics[width=\columnwidth]{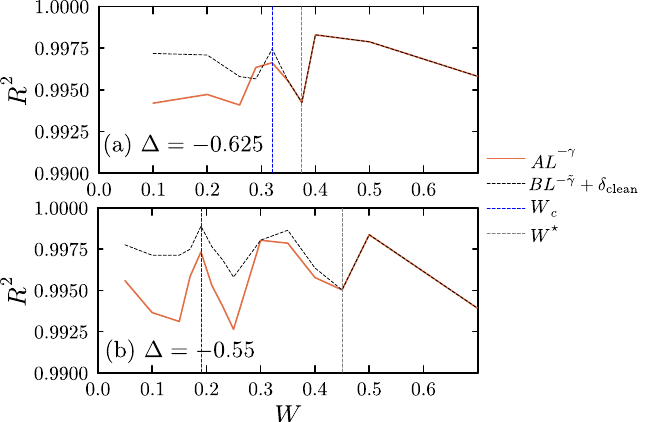}
    \caption{Comparison of two possible fits to the data in the Giamarchi-Schulz regime of the transition: a pure power law, and a fit like Eq.~\eqref{eq:deltafit} with fixed constant $\delta_{\infty} = \delta_{\rm clean} := 1/2 - m_{\rm clean}$. (a) $\Delta = -0.625$, close to the Giamarchi-Schulz to weak-link crossover, $\gamma_0 = 0.315$. (b) $\Delta = -0.55$, deeper in the Giamarchi-Schulz regime. The critical disorder strengths are from Ref.~\cite{doggen_weak_2017}. We purposefully do not include errors in this Figure for readability. However, they are of the order of the irregularities seen in the plot. }
    \label{fig:fitqualities}
\end{figure}

We now turn to the Giamarchi-Schulz regime of the SF-BG transition, see Fig.~\ref{fig:PhaseDiag}. We argue that the nature of the transition, together with the slow decay of the minimal deviation, makes it difficult to pinpoint the transition using extreme-magnetization statistics.
Nevertheless, we make a conjecture about the behavior of the minimal deviation at the transition based on (1)  numerical results for $\Delta = 0,\,-0.55,\,-0.625$, and (2) a continuity argument near $W_c \sim W^{\star}$ (where $W^{\star} = 1+ \Delta$ is the half-bandwidth of the clean problem, see Fig.~\ref{fig:PhaseDiag}). 

First, recall that, in the weak disorder regime, the SF is destroyed not by the effect of weak links systematically getting weaker, but rather by the proliferation of topological defects in the phase (quantum phase slips)~\cite{giamarchi_anderson_1988,giamarchi_localization_1987,giamarchi_quantum_2004}. Thus, we do not expect the local density to play a key role at the transition. Second, we saw that for $\gamma_{\rm eff} \lesssim 0.2$, finite-size effects are expected to play an important role and possibly lead to the false conclusion that $\gamma_{\rm eff} \rightarrow 0$. In practice, these two effects limit the use of extreme statistics to locate the Giamarchi-Schulz transition.

This is shown in Figs.~\ref{fig:gammaeff} and~\ref{fig:fitqualities}. The first figure shows that $\gamma_{\rm eff}$ seems to decrease with increasing system sizes for $W \lesssim W^{\star}$. However, this occurs for already small exponents, where we expect the finite-size effects to underestimate the power-law exponent. Relying on the observation that in the SF, $\delta_{\min}^{\typ} > \delta_{\rm clean}$ (see Appendix~\ref{app:distributions}), we attempt a different analysis, by performing a power-law fit with $BL^{\tilde{\gamma}} + \delta_{\rm clean}$, where $\tilde{\gamma}$ is free. Comparing the fit qualities in Fig.~\ref{fig:fitqualities} (for $\Delta = -0.55$ and for $\Delta = -0.625$), we see that the results are rather inconclusive. They suggest that the fit \emph{with} a constant is marginally better in the SF regime, and in part of the Bose glass phase for $W \lesssim W^{\star}$, but the large oscillations reflect large errors. 

Can we make a conjecture about the thermodynamic limit? 
Let us first focus on the transtion. Recall that the power-law exponent $\alpha$ controlling the distribution of the weak links $P(J) \sim J^{\alpha}$ diverges at the Giamarchi-Schulz transition~\cite{doggen_weak_2017}, i.e. the probability of very weak links vanishes, consistent with the transition being controlled by quantum phase slips. This implies that the weakest link scales with $J_{\rm min} \propto L^{-\frac{1}{1+\alpha}}$, with $\alpha$ diverging. For the same reason, we expect $\gamma$ to be zero along the Giamarchi-Schulz transition, and $\delta_{\infty} > 0$. A further argument supporting this conjecture is the observation that in the SF, the distributions seems systematically lower-bounded by the deviation in the clean case (see Appendix~\ref{app:distributions}). If this holds in the limit of large system sizes, then to have a continuity of $\delta_{\infty}$ transition point for $W= W^{\star}$, $\delta_{\infty}$ should become non-zero at the Giamarchi-Schulz transition. 

The question then becomes whether $\delta_{\infty} > 0$ or $\delta_{\infty} = 0$ in the Bose glass above the Giamarchi-Schulz transition. One possible scenario would be that $\delta_{\infty} > 0$ in the Bose glass below $W^{\star}$. This would suggest that, in this regime, the local density does not see a trace of the localization transition~\footnote{This would \emph{a priori} make sense given that the GS transition is driven by quantum phase slips.}, re-opening the possibility of two distinct Bose glass phases~\cite{giamarchi_localization_1987,ristivojevic_superfluid_2014,doggen_weak_2017}, here below and above the half-bandwidth. 

However, we observe from the analysis at $\Delta = 0$ and a verification at very weak interactions that the minimal deviation can go below $\delta_{\rm clean}$ on the localized side below the half-bandwidth, and that $\delta_{\infty} =0$ is compatible with the results at $W = 0.9 < W^{\star}$. This rather suggests a scenario where $\delta_{\infty} = 0$ in the Bose glass, but where strong finite-size effects hide this result.

\begin{figure}[t!]
    \includegraphics[width=0.9\columnwidth]{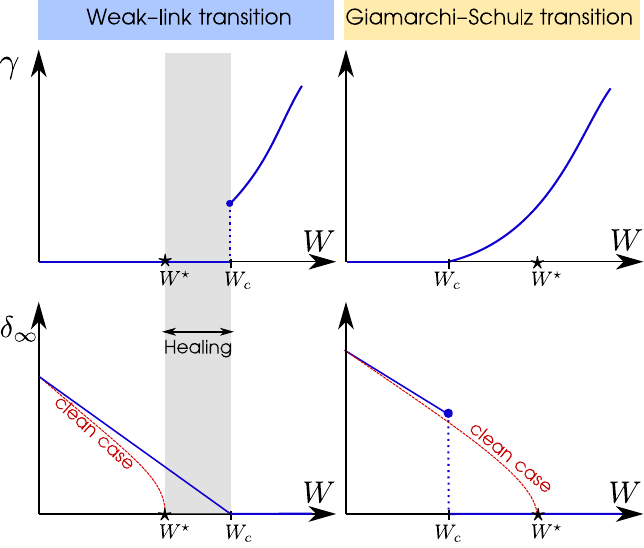}
    \caption{\label{fig:conjecture} Sketch of the conjectured behavior of the extreme value exponent $\gamma$ and limit minimal deviation $\delta_{\infty}$ in the weak-link respectively Giamarchi-Schulz regimes of the transition. Top panels: exponent $\gamma(W)$ as a function of the disorder strength $W$. Bottom panels: $\delta_{\infty}(W)$ (blue lines), compared to $\delta_{\rm clean} := 1/2 - m_{\rm clean}$ (dashed red). In all panels, $W^{\star} = 1+ \Delta$ (star-shaped symbol) indicates the half bandwidth, while $W_c$ (dash) indicates the critical disorder. Dotted blue lines indicate a jump. Full blue circles highlight the value \emph{at} the transition. The highlighted gray area indicates the region where the minimal deviation is ``healed''. }
\end{figure}

This is summarized in Fig.~\ref{fig:conjecture}: at the weak-link transition, the minimal deviation probes a proxy of the weak-links driving the transition and our numerics predicts that $\delta_{\infty}$ goes to zero at $W = W_c$, whereas in the Giamarchi-Schulz regime, we conjecture that $\delta_{\infty}$ jumps at the transition.

\section{Summary and outlook}
\label{sec:outlook}

In this work, we have investigated the use of a simple observable which is the extreme value of the magnetization (density) to capture a delocalization-localization transition in the ground state of the random-field XXZ chain.  This basic metric highlights the sharp contrast between the two regimes of the superfluid-Bose glass transition in this paradigmatic model. Indeed, we have demonstrated that in the strong disorder regime driven by weak-links, it successfully captures the location of the transition and the existence of a BKT-like scaling, with relatively modest system sizes. In the weak-disorder (Giamarchi-Schulz) regime, the extreme magnetization is essentially unaffected by the transition for available system sizes, due to extremely important finite-size effects and the completely different mechanism driving the transition. We nevertheless proposed a conjecture for the behavior of this observable in the infinite system size limit, where the chain breaking should be recovered in the Bose glass. In this sense, this part of the study may serve as a cautionary tale for an apparent absence of localization depending on the observable. Yet, our results are also reminiscent of discussion around the possible existence of two \emph{distinct} Bose glass phases~\cite{giamarchi_localization_1987,ristivojevic_superfluid_2014,doggen_weak_2017}: indeed, from the perspective of the local density on finite-size chains, the Bose glass phase looks quite different above and below the half-bandwidth. We also note that, if a transition exists between two distinct Bose-glass regimes, the proximity of a tricritical point could account for the observed $\nu_{\mathrm{loc}}\approx 0.67(3)$, which departs from the BKT value $\nu = 1/2$.

The rare events we have investigated are intrinsically linked to entanglement bottlenecks~\cite{dupont_from_eigenstate_2019}, but are experimentally easier to measure, as compared to the entanglement entropy, which is limited to very small systems~\cite{lukin_probing_2019,chiaro_direct_2022}. For instance, extreme magnetization may be probed in bulk material through NMR spectroscopy, which  can resolve inhomogeneous magnetic moment profiles~\cite{Tedoldi_NMR_1999,kramer_spatially_2013,zhou_quasiparticle_2017}. We also note (Appendix~\ref{app:DMRG}) that it may not be necessary to use the absolute \emph{most} polarized site in the chain: the distribution of highly polarized sites contains sufficient information.

The role of extreme densities in our work highlights related interesting questions in MBL and out-of equilibrium dynamics. In particular, their possible connection with weak-links, which are key players in the destruction of the SF phase, suggests a similar question at high energy: in the strongly discussed avalanche phenomenology~\cite{de_roeck_stability_2017,luitz_how_2017,thiery_many-body_2018,crowley_avalanche_2020,morningstar_avalanches_2022}, what is the role played by such bottlenecks? More precisely, it would be relevant to study how extremely polarized sites in an initial state respond to the presence of a thermal bubble in their vicinity. 

Finally, we note that our approach is conceptually simple in spirit, it may be helpful to better understand glassy aspects of the Bose glass~\cite{Dupuis_2019,Dupuis_2020,Daviet_Chaos_2021}, and could be easily generalized to other models, for instance quasiperiodic potentials~\cite{aubry_1980,cookmeyer_critical_2020,vongkovit_effective_2024}, long-range hopping models~\cite{Dupuis_2024} or other types of interactions or degrees-of-freedom~\cite{vishnu_phases_2025,bahovadinov_tomonaga-luttinger_2024}.

\acknowledgments
It is a pleasure to thank Nicolas Dupuis for numerous insightful discussions on this topic. We are also deeply indebted to Georges Bouzerar, Sylvain Capponi, Weitao Chen, Thierry Giamarchi and Nicolas Roch for enlightening discussions and comments. We are grateful to Maxime Dupont for sharing his C++ DMRG code at the beginning of the project. All calculations are performed with the ITensor library in Julia~\cite{ITensor,ITensor-r0.3}.  NL and GL are supported by the ANR research grant ManyBodyNet No. ANR-24- CE30- 5851. This work has been partly supported by the EUR grant NanoX No. ANR-17-EURE0009 in the framework of the ”Programme des Investissements d’Avenir”, is part of HQI initiative (www.hqi.fr) and is supported by France 2030 under the French National Research Agency award numbers ANR-22-PNCQ-0002 and ANR-23-PETQ-0002, and also benefited from the support of the Fondation Simone et Cino Del Duca. JC, GL, and SA acknowledge support from the Singapore Ministry of Education AcRF Tier 2 grant MOE-T2EP50222-0005. NC acknowledges LPT Toulouse (CNRS) for hospitality and the support from Delft Technology Fellowship. We acknowledge the use of HPC resources from CALMIP (grants 2023-P0677, and 2024-P0677) and GENCI (project A0150500225 and A0170500225).

\setcounter{section}{0}
\setcounter{secnumdepth}{3}
\setcounter{figure}{0}
\setcounter{equation}{0}
\renewcommand\thesection{S\arabic{section}}
\renewcommand\thefigure{S\arabic{figure}}
\renewcommand\theequation{S\arabic{equation}}

\appendix

\begin{center}
    \bfseries\Large Appendix
\end{center}

\section{Extreme magnetization in the XX chain ($\Delta = 0)$}
\label{app:AL}
This appendix provides several details related to Section~\ref{sec:noninteracting}.
We first recall some key ideas regarding the extreme magnetization in the  $\Delta = 0$ limit (non-interacting fermions, see Section~\ref{sec:noninteracting}) at high energy. For further detail, see the work of two of the present authors in Ref.~\cite{colbois_breaking_2023} and references therein. We then discuss new results highlighting how this picture changes in the ground state. All the presented numerical results in this section are extracted from exact diagonalization of large periodic chains (up to $L = 4096$), focusing on the $S^{z}_{\rm tot} = 0$ sector for all the same disorder realizations as used in Ref.~\cite{colbois_breaking_2023}. 
A key ingredient here will be the single particle localization length $\xi(W, \mathcal{E})$ (obtained from the Lyapunov exponent~\cite{Kramer_1993,Crisanti_1993,Comtet_2013}). The key point is that it depends both on the single particle energy $\mathcal{E}$ and on the disorder strength. Its most important features are (1) it is maximal in the middle of the band~\cite{Kappus_1981} and (2) it has weak tails at the band edges (where the localization length can be one order of magnitude smaller than in the middle of the band (see e.g. Ref.~\cite{colbois_breaking_2023}, Appendix S2), and (3) for $W \geq 2$, the localization length $\xi(W, \mathcal{E}) < 1\, \forall\, \mathcal{E}$. 

\begin{figure}
    \centering
    \includegraphics[width=0.5\linewidth]{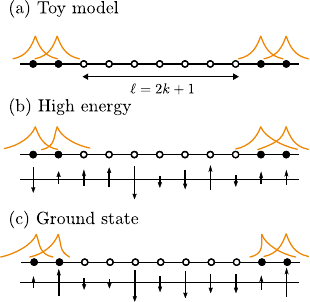}
    \caption{Argument for the relation between the minimal deviation and the localization length in the non-interacting case. Localized orbitals are sketched in orange; black and white dots correspond to the occupied (resp. empty) \emph{orbitals}, positioned at their localization center in real space; and arrows correspond to field configurations in real space. A very similar sketch would apply to a series of occupied orbitals surrounded by empty orbitals. (a) In the toy model, all orbitals are assumed to have the same localization length. The occupation (and thus the deviation to perfect polarization) in the center of a series of empty neighbouring orbitals is dominated by $\exp^{-\frac{\ell}{2\xi_{\rm toy}}}$. (b) In the XX chain in the middle of the many-body spectrum, the occupation of orbitals is essentially independent of the field realization, but orbitals are not necessarily symmetric and can have various localization lengths. (c) In the ground state at strong disorder, the occupation is directly related to the field configuration in real space. This results in particularly short localization lengths controlling the most polarized site. }
    \label{fig:toymodelsketch}
\end{figure}
\subsection{High energy: toy model and its validity}
We provide slightly more detail on the toy model briefly mentioned in Sec.~\ref{sec:spinfreezing}, which seeks to characterizes the minimal deviation decay in the free fermion case, and recall its range of validity. A detailed discussion can be found in Ref.~\cite{colbois_breaking_2023}. 

Recall that, in the Jordan-Wigner mapping~\cite{jordan_uber_1928}, the local magnetization $S^{z}_i$ relates to the local fermion density as $S^{z}_i = n_i - 1/2$, and in the free fermion case, the many-body wavefunction is a Slater determinant of the single-particle wavefunctions, with half-filling corresponding to the $S^z_{\rm tot} = 0$ sector. Therefore, focusing on the extreme magnetization is equivalent to focusing on the emptiest or on the most occupied site in the fermionic language. The density on one site is the sum of the contributions of all occupied orbitals on this site. 

The toy model describes the free fermions model in a random potential as a collection of localized orbitals (labeled by $m$), with two core simplifying assumptions : (1) orbitals are centered each on a different site $i_m$ and (2) they all have the \emph{same} localization length $\xi_{\rm toy}$ : $|\phi_m(i)|^2 = A_m \exp\left(- \frac{|i - i_m|}{\xi_{\rm toy}}\right)$. In this simplified picture, the minimal deviation occurs in the middle of the longest series of $\ell_{\max}$ neighboring orbitals in real space, which all have the same occupation (either full or empty): see sketch in Fig.~\ref{fig:toymodelsketch}(a). For even-length clusters, this yields
\be
\delta_{\min}^{\toy} = \exp^{-\frac{\ell_{\max}}{2\xi_{\rm toy}}}
\ee

In the large system size limit, the longest region occurs once, i.e. the probability that it starts on a given site is $\mathcal{P}(\ell_{\max}) \propto 1/L$. At the same time, at half-filling, the probability that $\ell$ consecutive sites correspond to occupied orbitals is $\mathcal{P}(\ell) \propto 2^{-\ell}$. Equating these two probabilities, we get $\ell_{\max} \propto \ln(L)/\ln(2)$. Thus, finally one gets Eq.~\eqref{eq:toy}, recalled here:
\be
\delta_{\min}^{\toy} \propto L^{-\frac{1}{2\xi_{\rm toy} \ln 2}}.
\label{eq:toyapp}
\ee
When considering strong enough disorder compared to the hopping term for fermions (or transverse coupling for the spin chain), the localization lengths become very short, such that most of the orbitals can be associated with a single site. It is also sufficiently sharply distributed.We therefore expect the toy model to hold in this regime. 

Thus, in the middle of the many-body spectrum, for large enough system sizes, both the location of the most polarized site and the value of the minimal deviation are essentially independent of the random potential (random field), since the orbitals selected to participate in the many-body wavefunction are occupied essentially irrespective of their energy~\footnote{See Ref.~\cite{colbois_breaking_2023} for a more precise statement.}, see Fig.~\ref{fig:toymodelsketch}(b). In this case, one can identify $\xi_{\rm toy}(W)$ with the Anderson localization length averaged over the density of states $\xi_{\rm AL}(W)$. In this context, the relation Eq.~\eqref{eq:toy} between the decay exponent and the localization length was validated through extensive numerical simulations at strong disorder in the Anderson-localized case at high energy~\cite{colbois_breaking_2023}.

The toy model also predicts that the minimal deviation distributions should show well-defined peaks at values of $\ln \delta^{\toy}_{\min}$ spaced approximately by $1/\xi_{\rm toy}$ \footnote{This is the spacing between peaks corresponding to length $\ell$ clusters of the same parity. Clusters of length $\ell = 2k$ actually give a peak located at a slightly smaller value than clusters of length $\ell = 2k-1$. }
In the XX chain, the distribution of the localization lengths effectively smoothes out the distributions of minimal deviations, and blurs the peaks~\cite{colbois_breaking_2023}. As a consequence, at intermediate disorder strengths where some localization lengths are large, the distributions do not show peaks. However, at strong disorder, when the localization length becomes very short compared to the lattice spacing, clear peaks appear, related to the integer length $\ell_{\max}$ of the cluster containing the most polarized site. This is obtained for $W \gtrsim 10$ in Ref.~\cite{colbois_breaking_2023}. 

\subsection{Ground state minimal deviation for $W > 1.2$}
We now argue and verify that at strong enough disorder in the XX chain, the decay of minimal deviation with increasing system sizes is faster in the ground state than at high energy. We build the argument using the longest cluster of neighbouring empty orbitals, but a similar argument holds for the longest cluster of occupied orbitals.

In the toy model picture, the minimal deviation occurs in the middle of the longest cluster of neighbouring empty orbitals. Therefore, as mentioned above, it is dominated by the localization length of the two occupied orbitals that occur at the edges of this cluster. Unlike high-energy states, the ground state corresponds to the occupation of the $L/2$ orbitals lowest in energy. First, this implies that half of the orbitals with the smallest localization lengths (which occur on band edges) are occupied. Second, at strong enough disorder, the occupation of orbitals is directly correlated to the field realization. As a result, the minimal deviation is now expected to occur in the middle of the longest series of fields that have the same sign, see Fig.~\ref{fig:toymodelsketch}(c). A simple perturbation theory argument shows that field differences between neighbouring sites control the shape of the localized orbitals. Because of the larger field difference at the boundary of the cluster, the two occupied orbitals at the end of the cluster must be strongly asymmetric and have a very short localization length in the direction of the cluster. Both effects conspire such that the minimal deviation in the ground state will typically be dominated by the shortest localization lengths, i.e. $\xi_{\rm toy} \sim \xi_{\min} (W) \ll \xi_{\rm AL}(W)$.  As a result, for the same disorder strength, the minimal deviation will decay faster with increasing system sizes in the ground state than at high energy. 

\begin{figure}
    \centering
    \includegraphics[width=0.95\columnwidth]{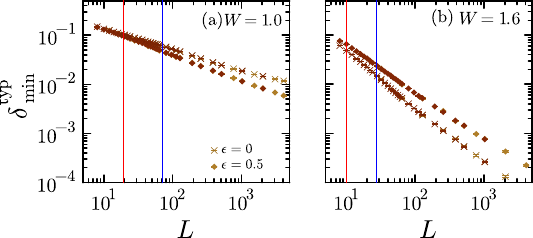}
    \caption{Typical minimal deviation decay for $\Delta = 0$ for two disorder strengths. The two colors correspond to simulations with different number of samples ($10^4$ in dark red and $10^3$ in yellow). Crosses show ground-state results ($\epsilon = 0$) and diamonds show high energy results. The vertical lines show two different scales related to the localization length: $L = 10 \xi$ in red and $\ln L = 10 \xi$ in blue.}
    \label{fig:ALdecay}
\end{figure}

\begin{figure}
    \centering
    \includegraphics[width=0.95\columnwidth]{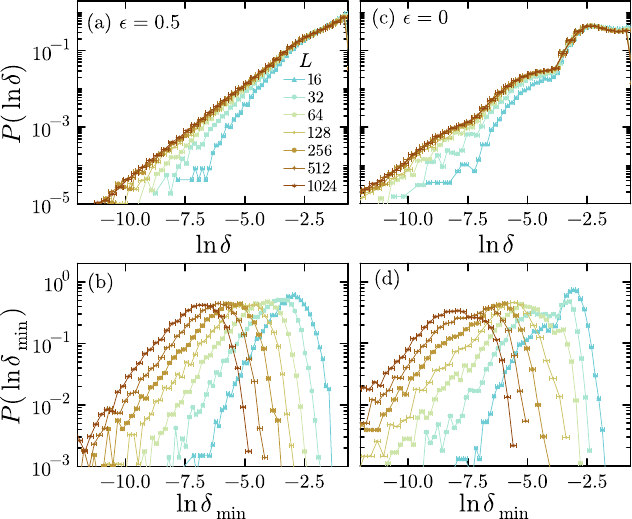}
    \caption{Distributions of the deviations and of the minimal deviations at $\Delta = 0$ for $W = 1.6$. (a) and (b) Distributions at high energy; (c) and (d) distributions for the same disorder realisations but in the ground state (half-filling). }
    \label{fig:distW1.6}
\end{figure}
In Fig.~\ref{fig:ALdecay} (focus on panel (b) for now), we show a comparison of the decay of the typical minimal deviation in the ground state ($\epsilon = 0$) v.s. in the middle of the spectrum ($\epsilon = 0.5)$ for $W = 1.6$, for the same disorder realizations. Although the disorder is not very strong, we clearly observe that the minimal deviation decays faster in the ground state than at high energy, in agreement with the above discussion. The fact that states with very short localization lengths are occupied has another important effect: the finite-$\ell$ effects in the distributions are already seen at much weaker disorder. This is very clear when comparing the distributions at high energy \emph{v.s.} in the ground state for $W = 1.6$ in Fig.~\ref{fig:distW1.6}. Indeed, at high energy, panel (a) shows an exponential tail for $\mathcal{P}(\ln(\delta))$ corresponding to a power-law tail for $\delta$, and as a result, panel (b) shows a simple ``travelling-wave'' behavior for the distributions of $\ln(\delta_{\min})$ except for the smallest system sizes (see Ref.~\cite{colbois_breaking_2023} for a detailed discussion). In contrast, in the ground state, panel (c) shows bumps (induced by finite-$\ell$ effects) appearing in the distributions of the deviations and minimal deviations. They result in irregular shapes for the distributions of the minimal deviation in panel (d). Through the toy model picture, the spacing between the main peaks gives an estimation for the minimum localization length $\xi \sim 1/2.5 = 0.4 $, in agreement with results obtained from the Lyapunov exponent which give $\xi_{\min}(W= 1.6) = 0.399$ (the average localization length from the same calculation gives $\xi_{\rm AL}(W=1.6) = 0.938$).

\subsection{Ground state vs high energy minimal deviation for weak disorder.}
For $W \leq 1.2$, the largest localization length becomes larger than 2 lattice spacings, and the toy model picture starts breaking down. 

We recall the main results at high energy~\cite{colbois_breaking_2023} for weak disorder (see also Figs.~\ref{fig:ALdecay}). The power-law decay of the minimal deviation keeps holding for $L \gtrsim 10 \xi_{\rm avg}$ (red vertical lines). However (data not shown here), the exponent becomes larger than $\frac{1}{2 \xi_{\rm AL} \ln 2}$. In Ref.~\cite{colbois_breaking_2023}, it was observed that $\delta_{\min}$ remains controlled by $2\xi_{\rm AL}$, but that the length ($\overline{\ell_{\rm cluster}}$) of the region surrounding the most polarized site and in which the magnetization doesn't change sign is larger. This is easily understood: since single particle orbitals are very delocalized, if a single one is occupied in a streak of empty orbitals, even the on-site occupation will remain smaller than 1/2. In practice, the exponent is therefore controlled by $\overline{\ell_{\rm cluster}} \approx \ell_{\rm sc}(W) \ln(L)/\ln(2)$, where $\ell_{\rm sc} \rightarrow 1$ at strong disorder and diverges at weak disorder (see Fig.~9 in~\cite{colbois_breaking_2023}.

\begin{figure}
    \centering
    \includegraphics[width=\columnwidth]{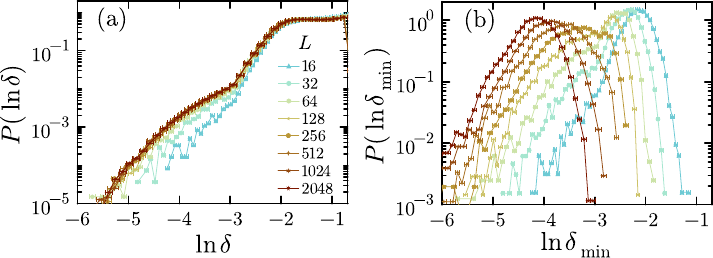}
    \caption{Distributions of the deviations (a) and of the minimal deviations (b) at $\Delta = 0$ for $W = 1$. }
    \label{fig:distW1AL}
\end{figure}
Another clear sign of the breakdown of the toy model is that for $W \leq 1.2$, the minimal deviation in the ground state decays \emph{slower} than at high energy (see for instance Fig.~\ref{fig:ALdecay}a). This could be explained by the increasingly large $ \gamma > \frac{1}{2 \xi_{\rm AL} \ln 2}$ at high energy, or by opposite effects in the ground state, or both. 
For not \emph{too} weak disorder, some features from the toy model still hold: for instance, at $W=1$, the peaks induced by finite-size clusters still have a spacing compatible with $\xi_{\rm min} \sim 0.5$ (Fig.~\ref{fig:distW1AL}).  However, both Fig.~\ref{fig:ALdecay}(a) and Fig.~\ref{fig:distW1AL}(a-b) clearly suggest that $W = 1$ gets close to the limit imposed by finite sizes. 

\subsection{Overview}

\begin{figure}
    \centering
    \includegraphics[width=\columnwidth]{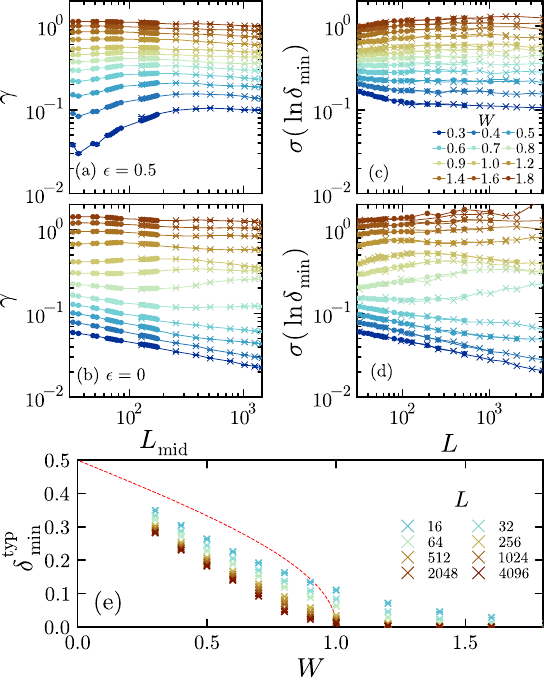}
    \caption{Overview of the distributions at high energy and in the ground state for the non-interacting case. (a)-(b) Effective exponent for sliding fits. (c)-(d) standard deviation of the distributions. In both cases, dots are data extracted from $10^4$ samples and crosses from $10^3$ samples. (e) Typical minimal deviation compared to the clean $\delta$.}
    \label{fig:overviewAL}
\end{figure}
Fig.~\ref{fig:overviewAL} provides an overview of the non-interacting case, by comparing the effective exponent $\gamma_{\rm eff}(L)$ as well as the standard deviation of the distributions of $\ln \delta_{\min}$ in the ground state and at high energy. As a consequence of the finite-$\ell$ effects which are more pronounced in the ground state, we observe a non-monotonous behavior $\gamma_{\rm eff}$ and $\sigma$ which is not present at high energy. We ascribe these slow oscillations to the finite-$\xi$ broad features present in the distributions, see Fig.~\ref{fig:distW1AL} for instance.

Despite these non-monotonous effects, sliding power-law fits capture fairly well that $\gamma_{\rm eff}(L) \rightarrow \gamma > 0$ and that $\delta_{\infty} =0$ for $W \geq 0.9$ (see panel Fig.~\ref{fig:overviewAL}(e)).
For weaker disorder, drawing conclusions is more difficult. We use $W = 0.7$ as a reference for an interesting effect. For fits on $L \leq 512$, Fig.~\ref{fig:overviewAL}(b) suggests that $\gamma_{\rm eff}(L) \rightarrow 0$. However, a study of larger system sizes clearly shows that $\gamma_{\rm eff}$ levels off around $\gamma \sim  0.121 \pm 0.03$, for $L \gtrsim 300$. For this disorder strength, the Anderson localization length is between $\xi_{\min} \sim 0.6$ and $\xi_{\max} \sim 6.6$ with average $\xi_{\rm AL}~\sim 3.4$: we see that this analysis seems to capture a constant exponent when $\ln L \sim \xi_{\max}$. In practice, focusing on the power-law exponent $\gamma$, this means that for the sizes available in the DMRG simulations, we can only capture correctly $\gamma \gtrsim 0.2$. For the non-interacting case, where we go up to $L = 2^{12}$, we capture the levelling off for $\gamma \sim 0.1$. For weaker disorders (which would correspond to an even smaller $\gamma$), we only see $\gamma_{\rm eff} \rightarrow 0$, and, based solely on our numerical data, we cannot conclude that $\delta_{\infty} = 0$ for $W < 0.6$. Given this observation, there are two key points:
(1) There is no fundamental difference between $W\sim 0.8$ and $W \sim 0.5$ in the Anderson insulator, and there is no reason to expect that $\delta^{\typ}_{\min}$ should behave fundamentally differently when considering sufficiently large sizes (up to scaling). Given that our largest size gives $\ln L \sim 8.3$ and that for $W \lesssim 0.6$ we have $\xi_{\rm AL}  \gtrsim 4.5$ and $\xi_{\rm max} \gtrsim 8.9$, it is reasonable to assume that our results are purely due to limitation in size.
(2) Crucially for the discussion in the main text, for $\Delta = 0$ we have that the typical minimal deviation for $W < 1$ is \emph{always smaller} than the $1/2-m_{\mathrm{clean}}$ (Fig.~\ref{fig:overviewAL}(e)). We verified that this remains true for weak interactions. 
This supports the idea that $\delta^{\typ}_{\min}(L) \rightarrow 0$ in the whole Bose glass phase, though it may be difficult to grasp when the localization length becomes large and the power-law exponent $\gamma$ becomes small.

\section{DMRG simulations and selection of extreme values with open boundary conditions}
\label{app:DMRG}
In the main text, we have focused on the extreme values of the magnetization obtained from DMRG simulations~\cite{ITensor,ITensor-r0.3,white_density_1992,schollwock_density-matrix_2011} in the interacting, random-field model of Eq.~\eqref{eq:ham}, on chains with open boundary conditions, and in the total magnetization sector $S^{z}_{\rm tot} = \sum_{i} S^{z}_i = 0$. The initialization is performed with a random MPS with bond dimension $\chi = 32$ in the $S^{z}_{\rm tot}= 0$ sector. 

In practice, all the results presented are obtained using density-matrix renormalization group (DMRG)~\cite{white_density_1992,schollwock_density-matrix_2011,ITensor,ITensor-r0.3} with conservation of the $U(1)$ charge, with up to $L = 128$ sites, maximal bond dimensions up to $\chi = 1024$, and requiring a convergence of $10^{-9}$ for the local magnetization, with 16 sweeps at most. Most of the time, results converged around 12 sweeps. For $\Delta = -0.75$, the results are averaged over 10'000 independent disorder realizations for $L < 64$, 5000 for $L\geq64$.  For $\Delta = -0.625$ and $\Delta = -0.55$, we used less samples. For $\Delta =-0.625$, we used more than $1500$ samples for $L \leq 68$ and $1000$ samples for $L  \geq 68$. For $\Delta = -0.55$, we consistently used 2000 samples or more. 

We now discuss boundary conditions (BC). Working with either open or fixed BC implies the presence of edge effects. Thus, it would be preferable from a purely physical perspective to use periodic BC in order to avoid these effects altogether. However, with periodic BC, DMRG requires a significantly larger bond dimension to reach similar precisions~\cite{schollwock_density-matrix_2011,pippan_efficient_2010}. If the goal is to minimize the required bond dimension (or, alternatively, to get the most out of a given bond dimension), fixed BC is preferable to open BC~\cite{Laflorencie_boundary_2006}; however, the two options are to either fix the boundary spins to be both aligned, which induces an overall magnetization in the remainder of the chain due to the conserved total magnetization, or to be anti-aligned, in which case the gain in entropy is limited. Therefore, we finally decided to work with open BC. 
 
As a result, we have to deal with non-negligible edge effects, as the two end spins experience half the coupling compared to the field. To minimize these effects, we select the most polarized site among the $L/2$ spins in the middle of the chain. If the selected site is on the edge of the selection region, we check whether it is a local minimum. We discard the rare samples where it is not, and check that we always keep more than 90\% of the samples.  We have performed two similar analysis to check that the results do not show a strong qualitative difference : (1) the same analysis with $3L/4$ sites in the middle of the chain  and (2) selecting instead the $n^{\rm th}$ most polarized site (for $n$ in $\{3,4,5\}$) and discarding the samples where this occurs on the edges. In  both cases, the main difference with the results shown in the main text is the presence of small effects depending on $mod(L,4)$ in the decay of the most polarized site, which makes the remainder of the analysis more challenging.

\section{Scaling}
\label{app:scaling}

\subsection{Disorder dependence of the amplitude and possible logarithmic corrections to scaling}

\begin{figure}
    \centering
    \includegraphics[width = 0.7\columnwidth]{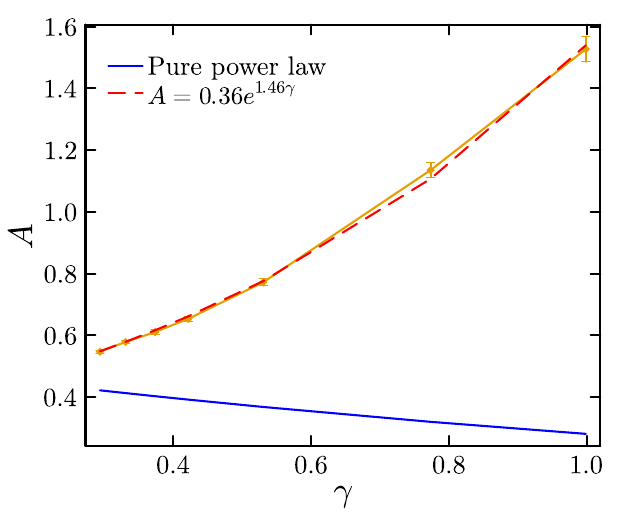}
    \caption{Relationship between $A$ and $\gamma$ for power-law fits for $\Delta = -0.75$ in the Bose glass, fit, and comparison with the expected dependence for pure power-law $\mathcal{P}(\delta)$ and $\gamma \ll 1/2$. }
    \label{fig:A}
\end{figure}

Fig.~\ref{fig:A} shows the presence of a disorder dependence for the amplitude $A(W)$ of the power-law Eq.~\eqref{eq:power-law} (in the interacting case). This highlights that there \emph{is} a disorder dependence of the amplitude controlling the minimal deviation.

Through the normalization of the full distribution of deviations $\mathcal{P}(\delta)$, the amplitude $D$ is related to the exponent $\gamma$. For instance, in the very crude approximation where the \emph{full distribution} is described by a power-law, 
\be
    \mathcal{P}(\delta) = D \delta^{\frac{1-\gamma}{\gamma}},
\ee
we have the relation
\be 
    D = \frac{1}{\gamma}2^{\frac{1}{\gamma}}
    \label{eq:amplitude}
\ee
This obviously is a gross underestimation in general. 

The existence of this relation implies the existence of a relation between the amplitude $A$ and the power-law exponent $\gamma$ for the minimal deviation decay. Indeed, a straightforward application of extreme value theory~\cite{majumdar_extreme_2020} to a distribution of deviations with a power-law tail gives, for the distribution of the  minimal deviation for a given system size~\cite{colbois_breaking_2023},
\be
    \mathcal{Q}_L(\delta_{\min}) = \eta DL\delta_{\min}^{\frac{1-\gamma}{\gamma}} \exp(-DL\gamma\delta_{\min}^{1/\gamma})
\ee
where $\eta = \frac{1}{1-e^{-DL\gamma2^{-1/\gamma}}}$ is a normalization factor close to one. 
In particular, the distribution of the \emph{logarithm} of the minimal deviation $u = \ln \delta_{\min}$ is
\be
    \mathcal{P}_L(\ln \delta_{\min}) = \eta D L \exp\left[\frac{u}{\gamma} - DL\gamma e^{u/\gamma}\right].
\ee
In Ref.~\cite{colbois_breaking_2023}, this expression  was used to estimate the typical minimal deviation as the most probable value. This is of course an approximation, and  in practice one should  perform the integral to compute the expectation value $\overline{ \ln \delta_{\min}}$ and extract the typical minimal deviation. The various divergent terms render an analytic derivation tricky in the regime where $\gamma \sim 1/2$ and we cannot make a reliable prediction for $A(\gamma)$ based on this analysis (see the difference between the predicted and observed dependence in Fig.~\ref{fig:A}).

Nevertheless, we note that in practice, the observed dependence $A(\gamma)$ in Fig.~\ref{fig:A} is compatible with an expression of the form 
\be 
A(\gamma) = a_1 \exp(a_2 \gamma)
\label{eq:simple}
\ee
This implies that we can collapse the data with an expression of the form (Ref.~\cite{laflorencie_chain_2020})
\begin{equation}
    \ln \frac{\delta_{\min}^{\typ}(W)}{\delta_{\min}^{\typ}(W_0)} = g(\ln L/\lambda) + \frac{1}{\ln L} h(\ln L / \lambda)
\end{equation}
where
\begin{equation}
    g(X) = B - X;\quad  h(X) = CX
\end{equation}
and
\begin{equation}
   \lambda = \frac{1}{\gamma-\gamma_0}.
\end{equation}
Indeed, this scaling gives
\begin{equation}
\ln \frac{\delta_{\min}^{\typ}(W)}{\delta_{\min}^{\typ}(W_0)} = B + C (\gamma-\gamma_0) - (\gamma-\gamma_0)\ln L.
\end{equation}
This is consistent with the disorder dependence of Eq.~\eqref{eq:simple}  when $B =0$ and $C = a_2$.

\subsection{Scaling procedure}
In Sec.~\ref{sec:BKT} we performed a systematic scaling analysis of Eq.~\ref{eq:scaling2} as follows~\cite{slevin_corrections_1999,rodriguez_multifractal_2011,laflorencie_chain_2020,garcia-mata_scaling_2017}. 

In order to fit $L^{-\gamma_c} F$ to the data, we fix $W_c$ and Talyor-expand $F$ around $W \gtrsim W_c$ as
\be
    F((\ln L)^{\frac{1}{\nu_{\rm loc}}}) = \sum_{k=0}^{n} p_k  \rho^{k} (\ln L)^{\frac{k}{\nu_{\rm loc}}}
\label{eq:scalingfit}
\ee
to order $n = 4$, and also Taylor-expand $\rho$ to order $m=3$:
\be
    \rho(W-W_c) = (W-W_c) + \sum_{j= 2}^{m} x_j (W -W_c)^{j}.
\ee

The total set of parameters to be determined is $\{x_2, x_3, p_0, \dots p_4, \nu_{\rm loc}, \gamma_c\}$, i.e. $N_p = 9$.  We therefore have $N_{\rm{dof}} = N_d - N_p$ degrees-of-freedom for the goodness-of-fit evaluation, with $N_d$ the number of data for $W > W_c$ and any size.  We perform this scaling analysis systematically for different values of $W_c$ and show the results in Fig.~\ref{fig:Delta-075collapse}. A scaling with logartithmic corrections follows the same strategy and does not yield very different results, although the fit is less stable.

\section{More on the extreme values distributions in the interacting case}
\label{app:distributions}
In this Appendix, we present further results on the extreme statistics in the interacting case. We first discuss the strong disorder growth of the effective exponent $\gamma$ as well as the presence of effects similar to those observed in the non-interacting case deep in the Bose glass. We then illustrate the self-averaging of $\delta_{\min}$ in the SF, and finally comment on the evolution of the variance of the distributions with increasing sizes.

\begin{figure}
    \includegraphics[width=0.7\columnwidth]{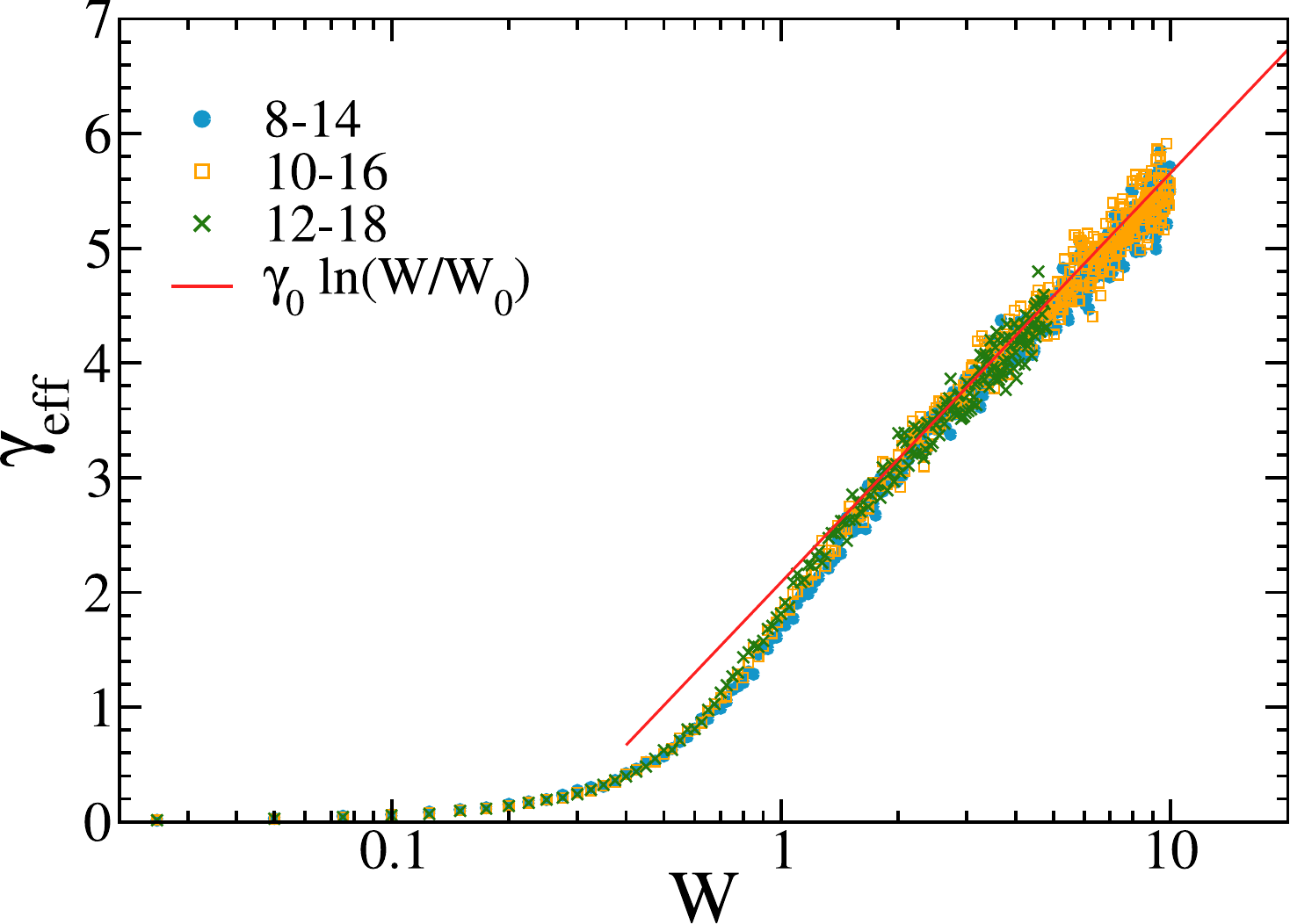}
    \caption{\label{fig:gammastrong} Disorder dependence of the effective decay exponent $\gamma_{\mathrm{eff}}$ obtained from sliding fits to ED results on 7500 independent disorder samples per disorder, at $\Delta = -0.75$, fitted is with $\gamma_0 = 1.55(5)$ and $W_0 = 0.26(1)$, in agreement within errors with $W^{\star} = 0.25$. Note the strong fluctuations due to the strong disorder regime deep in the Bose glass. }
\end{figure}

We first discuss the effective exponent deep in the Bose glass phase at $\Delta = -0.75$, Fig.~\ref{fig:gammastrong}. For such strong disorders ($W  \gg 1$, the highly localized regime is not amenable to reliable DMRG simulations. Instead, we rely on results from exact diagonalization, limited to small system sizes.  Despite the strong fluctuations induced by the disorder, which dominate over finite-size effects here, we observe a clear logarithmic dependence of $\gamma_{\rm eff}$ on the disorder.

\begin{figure}
    \includegraphics[width=0.7\columnwidth]{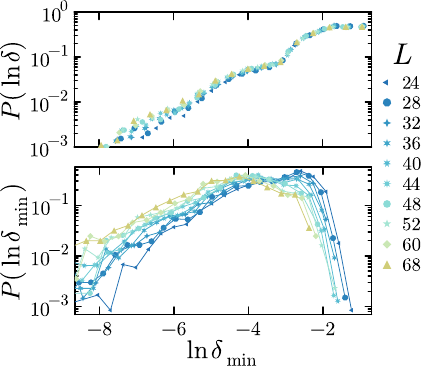}
    \caption{\label{fig:distW1Interacting} Distributions of the deviations (a) and the minimal deviations (b) for $\Delta = -0.625, W = 1$, obtained from DMRG simulations for 1500 independent samples. }
\end{figure}

\begin{figure}
    \centering
    \includegraphics[width=0.7\columnwidth]{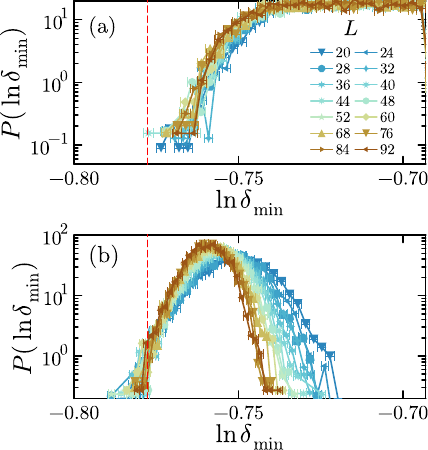}
    \caption{\label{fig:distW1InteractingSF}  Distributions of the deviations (a) and the minimal deviations (b) for $\Delta = -0.55, W = 0.05$, obtained from DMRG simulations for 2500 independent samples ($L\geq 36$) and more than 3000 samples ($L < 36$). The dashed red line shows the clean value for the deviation.}
    \label{fig:my_label}
\end{figure}
Regarding the distributions of the deviations and the minimal deviations, we note that the sub-structure induced by small localization lengths and seen in the non-interacting case (Fig.~\ref{fig:distW1.6}) are also present in the interacting case (Fig.~\ref{fig:distW1Interacting}) for strong enough disorder. They are much less pronounced than in the non-interacting case. In particular, the top panel also clearly highlight the exponential tail in the distribution of $\ln{\delta}$ corresponding to the power-law tails of $\delta$: $\mathcal{P}(\ln \delta ) \sim \exp[\frac{1}{\gamma} \ln \delta ]$. The distributions of the minimal deviations show a slight broadening (see the standard deviation below for a clearer picture). In contrast, in Fig.~\ref{fig:distW1InteractingSF}, we illustrate the distributions of the deviations and of the minimal deviations deep in the superfluid phase. The distributions of the minimal deviations are clearly self averaging. Crucially for our discussion in the main text, both distributions are approximately lower-bounded by the "clean" value $\delta_{\rm clean} := 1/2-m_{\rm clean}$. This is the "worst-case scenario" among all the studied cases, in that there is a very small probability to find a minimal deviation lower than the clean value for the smaller system sizes. It seems to reduce with increasing system sizes. 

\begin{figure}
     \centering
    \includegraphics[width=\columnwidth]{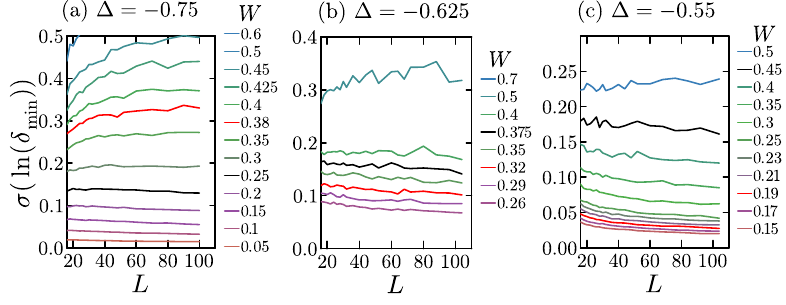}
    \caption{\label{fig:Dev}  Standard deviations of the distributions of $\ln \delta_{\mathrm{min}}$ (a) near the weak-link transition and (b)-(c) near the Giamarchi-Schulz regime. In all three cases, the critical disorder is shown in red, and the disorder corresponding to the bandwidth is shown in black. }
\end{figure}

To get an overview, in Fig.~\ref{fig:Dev} we compare the standard deviation of the distributions. In the two last panels of Fig.~\ref{fig:Dev}, we show the standard deviation of the distributions as a function of the disorder strength for varying system sizes and for $\Delta = -0.625, -0.55$. Both figures clearly show that for small sizes the distributions have a roughly constant variance near the bandwidth (suggesting self-similarity). In this regime, the disorder corresponding to the bandwidth is in the Bose glass phase. Closer to the SF-BG transition, the distributions are self-averaging, whereas they show a broadening deeper in the Bose glass. In contrast, for $\Delta= -0.75$ (panel (a)), a certain broadening already occurs in the superfluid phase and at the transition.

\newpage
\bibliography{sfbg}
\end{document}